\documentclass[preprints,article,accept,moreauthors,pdftex,10pt,a4paper]{Definitions/mdpi}
\firstpage{1} 
\pubvolume{xx}
\issuenum{1}
\articlenumber{1}
\pubyear{2018}
\copyrightyear{2018}
\externaleditor{Academic Editor: name}
\history{Received: date; Accepted: date; Published: date}

\pdfoutput=1

\renewcommand{\vec}[1]{\bm{\mathrm{#1}}}

\def \F{\vec{F}}

\def \x{\vec{x}}

\def\U {{{\bf u}}}
\def\F {{{\bf F}}}

\def\div {{\nabla \cdot}}

\usepackage{amssymb}
\usepackage{caption}
\usepackage{url}
\usepackage{amsmath}
\usepackage{float}
\usepackage{caption}
\usepackage{color}
\usepackage{graphicx}
\usepackage{rotating}
\usepackage{lscape}

\Title{Naut your everyday jellyfish model: Exploring how tentacles and oral arms impact locomotion}

\Author{Jason G. Miles$^{1}$ and Nicholas A. Battista $^{1}$*}

\AuthorNames{Jason G. Miles and Nicholas A. Battista}

\address{%
$^{1}$ \quad Dept. of Mathematics and Statistics, 2000 Pennington Road, The College of New Jersey, Ewing Township, NJ 08628}

\corres{battistn@tcnj.edu; Tel.: +1-609-771-2445}

\abstract{Jellyfish - majestic, energy efficient, and one of the oldest species that inhabits the oceans. It is perhaps the second item, their efficiency, that has captivated  scientists to investigate their locomotive behavior for decades. Yet, no one has specifically explored the role that their tentacles and oral arms may have on their potential swimming performance, arguably the very features that give jellyfish their beauty while instilling fear into their prey (and beach-goers). We perform comparative \textit{in silico} experiments to study how tentacle/oral arm number, length, placement, and density affect forward swimming speeds, cost of transport, and fluid mixing. An open source implementation of the immersed boundary method was used (\textit{IB2d}) to solve the fully coupled fluid-structure interaction problem of an idealized flexible jellyfish bell with poroelastic tentacles/oral arms in a viscous, incompressible fluid. Overall tentacles/oral arms inhibit forward swimming speeds, by appearing to suppress vortex formation. Non-linear relationships between length and fluid scale (Reynolds Number) as well as tentacle/oral arm number, density, and placement are observed, illustrating that small changes in morphology could result in significant decreases in swimming speeds, in some cases by downwards of 400\%  between cases with to without tentacles/oral arms.}


\begin{document}


%
%

%
%
%
%
%
%

\section{Introduction}

\begin{figure}[H]
\centering
\includegraphics[width=0.6\textwidth]{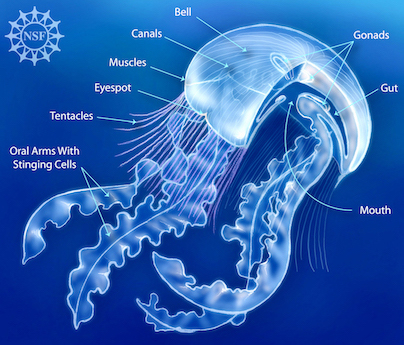}
\caption{Anatomy of a ``True" Jellyfish (class Scyphozoa). Courtesy of the National Science Foundation \cite{NSF:Anatomy}.}
\label{fig:Anatomy}
\end{figure}

A notably distinct (and arguably most beautiful) feature of a jellyfish are its tentacles and oral arms. Unfortunately these appendages are what ruin many summer days at the beach, as they're chalk full of their stinging cells, called nematocysts \cite{Higgens:2008,Beckmann:2012}. The movement of these stinging cells are one of the fastest movements in the animal kingdom, with discharges as fast as 700 nanoseconds \cite{Nuchter:2006}. Jellyfish use these cells to subdue their prey for sustenance \cite{Strychalski:2018}, hence their tentacles and oral arms play a vital role in survival. Figure \ref{fig:Anatomy} illustrate the anatomy of a ``true" jellyfish's tenatcles and oral arms.

\begin{figure}[H]
\centering
\includegraphics[width=0.999\textwidth]{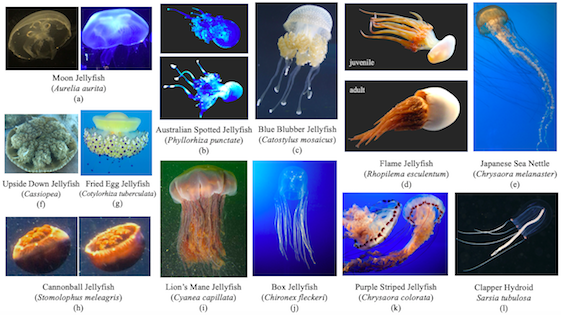}
\caption{Illustrating the diversity of tentacles/oral arms among different jellyfish species, including: (a) Moon jellyfish courtesy of the Two Oceans Aquarium \cite{TwoOceansAquarium} (left) and Audubon Aquarium of the Americas \cite{NOLAAquarium} (right) (b) Australian Spotted Jellyfish courtesy of the Aquarium of Niagara \cite{NiagaraAquarium} (c) Blue Blubber Jellyfish courtesy of H. Steiger \cite{Steigler:BlueBlubber}, (d) Flame Jellyfish courtesy of B. Abbott (juvenile, top) \cite{Abbott:JFJ} and the Osaka Aquarium Kaiyukan (adult, bottom) \cite{OsakaAquarium:FJ} , (e) Japanese Sea Nettle courtesy of the National Aquarium (USA) \cite{BaltimoreAquarium:JSN}, (f) Upside Down Jellyfish courtesy of \cite{KayLargoLab} (g) Fried Egg Jellyfish courtesy of \textit{Fredski2013} (left) \cite{Fredski2013} and A. Sontuoso (right) \cite{Sontuoso:FEJ}, (h) Cannonball Jellyfish courtesy of the National Aquarium (USA) \cite{BaltimoreAquarium:CB}, (i) Lion's Mane Jellyfish courtesy of D. Hershman \cite{Hershman:Liosnmane}, (j) Sea Wasp courtesy of the Port of Nagoya Public Aquarium \cite{BoxJelly:CC}, (k) Purple Striped Jellyfish courtesy of the Monterey Bay Aquarium \cite{MontereyBayAquarium:PSJ}, and (l) Clapper Hydroid courtesy of A. Semenov \cite{Semenov:Sarsia}. }
\label{fig:SpeciesDiversity}
\end{figure}

A ``true jellyfish" is one of a specific class of jellyfish - Scyphozoa. Scyphozoans tend to be the jellyfish that are familiar to aquarium-goers, identifiable by the cup shape of their bell. Another class of Medusozoa are Cubozoa (box jellyfish), denoted by their cube-shaped medusae. Both of these jellyfish classes are known to ruin beach-goers relaxed, fun day at beaches around the world \cite{Cegolon:2013}. There are notable visual differences in tentacle and oral arm morphology among different classes and species of jellyfish in general, see Figure \ref{fig:SpeciesDiversity}. Figure \ref{fig:SpeciesDiversity} presents 12 different jellyfish species illustrating different tentacle/oral arm morphologies- different numbers, lengths, configurations, and densities. Table \ref{table:morphology} gives specific morphological data pertaining to the species shown in Figure \ref{fig:SpeciesDiversity}. These jellyfish range on the smaller size, like the Clapper Hydroid (\textit{Sarsia tubulosa}), which has a bell diameter $\sim0.5\ cm$ with tentacles of length $\sim3-4\ cm$ \cite{Wrobel:2003,Katijai:2015}, to the Lion's Mane Jellyfish (\textit{Cyanea capillata}), which can have a bell of diameter $2.5\ m$ and tentacles/oral arms that can grow to be $36\ m$ long \cite{NationalAquarium:LionsMane}! Note that there are many more jellyfish species than those discussed above with different morphologies, like the Immortal Jellyfish (\textit{Turritopsis dohrnii}) or Blue Jellyfish (\textit{Cyanea lamarckii}).

\begin{landscape}

\begin{table}
\begin{center}
\begin{tabular}{| c | c | c | c |}
    \hline
    Name  & Scientific Name    & Max. Bell Diameter (cm)  & Tentacle/Oral Arm Length (cm) \\ \hline
    Moon Jellyfish      & \textit{Aurelia aurita}          & 38   &  $7.6$  \\ \hline
    Australian Spotted Jellyfish & \textit{Phyllorhiza punctata} & 60  &  $\gtrsim60$  \\ \hline
    Blue Blubber Jellyfish & \textit{Catostylus mosaicus}   & 45  &  $\sim45$  \\ \hline
    Flame Jellyfish       & \textit{Rhopilema esculentum}   & 70 &  $\gtrsim70$ \\ \hline
    Japanese Sea Nettle   & \textit{Chrysaora melanaster }  & 60 &  $300$ \\ \hline
    Upside Down Jellyfish &\textit{Cassiopea}              & 36  &  $36$  \\ \hline
    Fried Egg Jellyfish   &\textit{Cotylorhiza tuberculata} & 40 &  $\gtrsim40$ \\ \hline
    Cannonball Jellyfish  &\textit{Stomolophus meleagris}   & 25 &  $\gtrsim25$  \\ \hline
    Lion's Mane Jellyfish &\textit{Cyanea capillata}       & 250 &  $3600$  \\ \hline
    Sea Wasp (Box Jellyfish) & \textit{Chironex fleckeri}       &  $30$  & $300$  \\ \hline
    Purple Striped Jellyfish &\textit{Chrysaora colorata} & 70   &  $800$ \\ \hline
    Clapper Hydroid   & \textit{Sarsia tubulosa}       & 0.5  &  $3-4$ \\ \hline
    \hline
    \end{tabular}
    \end{center}
\end{table}

\begin{table}
\begin{center}
\begin{tabular}{| c | c | c |}
    \hline
    Name  & Range & References\\ \hline
    Moon Jellyfish & Slong the East \& West Coast, Europe, Japan and the Gulf of Mexico & \cite{MontereyBay:Moon} \\ \hline
    Australian Spotted Jellyfish & Western Pacific (From Australia to Japan) & \cite{Haddad:2006,Boon:Masters} \\ \hline
    Blue Blubber Jellyfish &Along the east and north coasts of Australia & \cite{MontereyBay:Blubber}  \\ \hline
    Flame Jellyfish &Warm temperate waters of the Pacific Ocean &\cite{Pitt:2009,Boon:Masters} \\ \hline
    Japanese Sea Nettle &Northern Pacific Ocean \& adjacent parts of the Arctic Ocean & \cite{ArcticOceanDiversity:SeaNettle} \\ \hline
    Upside Down Jellyfish & Western Atlantic, including the Gulf of Mexico, Bermuda and the Caribbean &\cite{GeorgiaAquarium:UpsideDown} \\ \hline
    Fried Egg Jellyfish & Mediterranean Sea, coastal lagoons &\cite{Kikinger:1992} \\ \hline
    Cannonball Jellyfish &Pacific Ocean and the mid-west Atlantic Ocean  &\cite{Griffin:CBJ} \\ \hline
    Lion's Mane Jellyfish & Cold waters of the Arctic, Northern Atlantic, Northern Pacific &\cite{NationalAquarium:LionsMane,McClain:2015} \\ \hline
    Sea Wasp & Australia and Indo-West Pacific Ocean &\cite{Barnes:1965,Lewis:2009} \\ \hline
    Purple Striped Jellyfish & Eastern Pacific Ocean primarily off the coast of California &\cite{GeorgiaAquarium:PSJ} \\ \hline
    Clapper Hydroid & Central California to the Bering Sea &\cite{Wrobel:2003,Katijai:2015} \\ 
    \hline
    \end{tabular}
    \caption{Morphological properties and range of the various jellyfish species shown in Figure \ref{fig:SpeciesDiversity}.}
    \label{table:morphology}
    \end{center}
\end{table}
\end{landscape}

The large diversity of jellyfish tentacle/oral arm morphology, leads to a variety of different modes of predation among different jellyfish species. For example, some jellyfish actively hunt for food, like box jellyfish \cite{Kinsey:1988}, while others are more passive and opportunist predators, hoping that their food (prey) gets caught in their tentacles/oral arms, like a Lion's Mane jellyfish \cite{Linnaeus:1758,Powell:2018}. Some species even use a combination of photosynthesis and passive filter-feeding for sustenance, e.g., the upside-down jellyfish \cite{Hamlet:2012,Santhanakrishnan:2012}. A jellyfish's foraging strategy may depend on its swimming performance potential, which may be limited by its morphology, not only by its bell but its tentacles and oral arms. If it is an effective/efficient swimmer, it display a more active predation strategy, while if it is not, it might display more passive, opportunistic foraging characteristics.

Some jellyfish, such as the Lion's Mane Jellyfish (\textit{Cyanea capillata}), predominantly move horizontally with the direction of prevailing currents \cite{BastianThesis:2011}. They also exhibit a range of vertical migrations depending on tidal stage \cite{Bastian:Poster,Moriarty:2012}. Costello et al. 1998 \cite{Costello:1998} observed Lion's Mane jellyfish sometimes active swimming upwards towards the surface for extended periods of time, to whom once at the surface would cease swimming and passively sink towards the bottom. Other times, they would reorient themselves and actively swim towards the bottom \cite{Costello:1998}. Using an acoustic tagging system, Bastian et al. \cite{BastianThesis:2011,Bastian:Poster} quantified their vertical speeds as approximately $3.75\ bell\ diameters/min$, or $0.06\ bell\ diameters/s$. Moreover, Moriarty et al. 2012 \cite{Moriarty:2012} found that Lion's Mane (and Fried Egg jellyfish) active swimming speeds varied during the tidal stage, with deviations of approximately 95\% from their maximal swimming speeds during the flooding stage. Although these jellyfish are known to be opportunist predators, they are effective carnivores \cite{Purcell:2001,Purcell:2003,Crawford:2016}; they are not fast swimmers (see above), but rather their lengthy, numerous dense tentacles compensate as an appropriate foraging advantage.

Other jellyfish species' migratory patterns are also influenced by the tidal stage, such as the sea wasp (\textit{Chironex fleckeri}), one kind of box jellyfish \cite{Gordon:2009}. Large box jellyfish tend to swim against the current \cite{Kinsey:1986}, while smaller individuals swim in either direction \cite{Rifkin:1996,Gordon:2014}. Box jellyfish have been observed achieving swimming speeds of up to $\sim1-2.5\ bodylengths\ per\ second$ and moreover that smaller individuals were observed swimming faster \cite{Colin:2013}. Compared to the Lion's Mane jellyfish, box jellyfish have a much more stiff bell, which allows them to achieve more rapid swimming \cite{Shorten:2005}. The effects of varying bell stiffness on propulsion in medusae have been quantified in computational models \cite{Hoover:2015,Hoover:2017}. Box jellyfish are able use this to their advantage when foraging, in tandem with their vision to hunt prey and relocate to densely populated prey environments \cite{Courtney:2015,Garm:2016}. As previously shown in Figure \ref{fig:SpeciesDiversity}, Lion's Mane and box jellyfish have distinct morphological differences, not only in their bell size and shape, but also their tentacle number, length, and density. Their bell kinematics and swimming behavior and performance are also different \cite{Grzimek:1972,Shorten:2005}, which supports the notion of that their foraging strategies depend on its morphology and potential swimming ability.

There are jellyfish that do a mixture of active and passive hunting. One such species is the \textit{Sarsia tubulosa}, who are ambush predators. They wait for its unsuspecting prey to come into its capture range, near its 4 tentacles, while it floats almost motionless in the water. They will actively swim to different locations to forage for food, but do so in irregular swim bouts, even in aquaria \cite{Leonard:1978}. Moreover, individual \textit{Sarsia} are known to each use a unique set of contraction frequencies while swimming \cite{Leonard:1980b}. Katija et al. 2015 \cite{Katijai:2015} found that \textit{Sarsia tubulosa} tend to swim about $2\ bodylengths\ per\ contraction$, with a range between \~[1,3.5] bodylengths per contraction. Similar results were obtained previously by Colin et al. 2002 \cite{Colin:2002} and Sahin et al. 2009 \cite{Sahin:2009}. Hoover et al. 2015 \cite{Hoover:2015} and Miles et al. 2019 \cite{Miles:2019} used an idealized jellyfish fluid-structure interaction computational model similar to the \textit{Sarsia tubulosa} and predicted swimming speeds within that range for biologically relevant $Re$ and quantified how different contraction frequencies may affect forward swimming performance. However, neither considered any tentacles/oral arms in their model. 

Many computational scientists have developed sophisticated computational models of jellyfish that produce forward propulsion \cite{Sahin:2009,Dular:2009,Wilson:2009,Hershlag:2011,Alben:2013,Yuan:2014,Hoover:2015,Hoover:2017,Hoover:2019,Miles:2019,Pallasdies:2019} and have compared swimming performance over a large mechanospace, including bell stiffness and flexibility, muscular contraction strength and amplitude, and contraction frequencies. While many \textit{in situ} and laboratory studies have been performed that quantify bell kinematics, velocity, acceleration, feeding (predation), and/or vortex wake structure \cite{Costello:1998,Colin:2002,Dabiri:2005,Shorten:2005,Dabiri:2005b,Dabiri:2006,Bajcar:2008,Peng:2009,Colin:2013,Gemmell:2013,Gemmell:2014,Katijai:2015,Courtney:2015,Gemmell:2015,Costello:2019,Kim:2019}, computational studies are becoming increasingly more of an attractive alternative and/or complement for scientists, as they is easier (e.g., more cost and time efficient) to do widespread parameter studies. Yet, with all of the data available no one has specifically explored the role that tentacles/oral arms may have on jellyfish locomotion. The only study the authors are aware of was performed by K. Katija in 2015 \cite{Katija:2015}, where they observed that an Australian Spotted Jellyfish (\textit{Phyllorhiza punctata}) swam 360\% faster without its oral arms; however, any further investigations have yet to have been performed. 

In this paper, we will use an open-source implementation of the immersed boundary method, \textit{IB2d} \cite{Battista:2015,BattistaIB2d:2017,BattistaIB2d:2018}, to study the swimming performance of a $2D$ idealized jellyfish with tentacles/oral arms within a fully-coupled fluid-structure interaction framework. We expand upon the work of Hoover et al. 2015 \cite{Hoover:2015} and Miles et al. 2019 \cite{Miles:2019}, using their geometric bell morphology as a base model, but with the addition of poroelastic tentacles/oral arms. In particular studied how the addition of tentacles/oral arms will effect its potential forward swimming ability, across numerous fluid scales, given by the Reynolds Number, $Re$, and lengths, number, density, and placement of tentacles/oral arms. We also show that intricate relationships exist between vortex interactions, wake structure, fluid mixing, and ultimately their forward propulsive speed, highlighting that there appears to be no one-to-one connection between swimming efficiency and wake structure \cite{Smits:2019,Floryan:2019}. 

In addition, we offer the science community the first open-source jellyfish locomotion model with poroelastic tentacles/oral arms in a fluid-structure interaction framework. It can be found at \url{https://github.com/nickabattista/IB2d/tree/master/matIB2d/Examples/Example_Jellyfish_Swimming/Tentacle_Jelly}. It is worthwhile to comment that this is a generalized study of how tentacles/oral arms may effect forward swimming performance in jellyfish and are not modeling one particular species. 

\section{Mathematical Methods}
\label{sec:methods}


To model the interactions between a flexible jellyfish bell with tentacles/oral arms and a viscous, incompressible fluid, computational methods were used. Our goal was to explore forward swimming performance over a wide parameter space, including fluid scale ($Re$) and tentacle/oral arm length, number, density, and placement. While fluid scale effects have been previously studied for jellyfish locomotion \cite{Hershlag:2011,Hoover:2017,Miles:2019}, how tentacles/oral arms may impact forward swimming has yet to been investigated.

The mathematical framework used is in the vein of fluid-structure interaction systems (FSI), which couples the motion of a deformable object and the fluid to which it is immersed. The first robust numerical method developed to solve such problems was developed by Charles Peskin in the 1970s \cite{Peskin:1972,Peskin:1977}. It is called the \textit{immersed boundary method} (IB) \cite{Peskin:2002}. Due to its elegance, the immersed boundary method has been continually used and improved upon \cite{Fauci:1993,Lai:2000,Cortez:2000,Griffith:2005,Mittal:2005,Griffith:2007,BGriffithIBAMR,Griffith:IBFE,Huang:2019}. It is still a leading numerical framework for studying problems in FSI due to its robustness \cite{BattistaIB2d:2017,BattistaIB2d:2018} that extends well beyond the physiological (or biological) applications it was originally developed for, and has permeated across all fields of engineering \cite{Mittal:2005,BattistaIB2d:2018,Huang:2019} . 

The IB has been successfully applied to numerous problems including cardiac fluid dynamics \cite{Miller:2011,Griffith:2012b,Battista:2016a,Battista:2017,Battista:2018}, aquatic locomotion \cite{Bhalla:2013a,Bhalla:2013b,Hamlet:2015}, insect flight \cite{Miller:2004,Miller:2009,SJones:2015}, and even dating and relationships \cite{BattistaIB2d:2017}. Additional details on the IB method can be found in the Appendix \ref{IB_Appendix}.

We will now dive into the details regarding our jellyfish model's implementation into the \textit{IB2d} framework, e.g., the computational geometry, geometric and fluid parameters, and modeling assumptions. Our model is based upon the $2D$ jellyfish locomotion model of Hoover et al. 2015 \cite{Hoover:2015}, which originally did not include tentacles/oral arms. It was originally implemented in the IB software called IBAMR \cite{BGriffithIBAMR}, which is parallelized IB software with adaptive mesh refinement \cite{MJBerger84,Roma:1999,Griffith:2007}.

%
%

\subsection{Computational Parameters}
\label{sec:parameters}

In this study, we use the frequency-based Reynolds number, $Re_f$, to describe the locomotive processes of \textit{prolate} jellyfish with tentacles/oral arms. The characteristic length, $D_{jelly}$, is set to the bell diameter at rest and the characteristic frequency, $f_{jelly}$, is set to the contraction (stroke) frequency. Therefore our characteristic velocity scale is set to $V_{jelly}=f_{jelly}D_{jelly}$, as in Eq.(\ref{eq:Re_study}),
\begin{equation}
    \label{eq:Re_study} Re = \frac{\rho f_{jelly} D_{jelly}^2 }{\mu}.
\end{equation}
Fluid parameters (density and viscosity) can be found in Table \ref{table:num_param}. Note that for our study in Section \ref{results:sec_Re}, we only vary the viscosity, $\mu$, which effectively changes the Reynolds Number, $Re$ (see Eq.(\ref{eq:Re_study})), and the number, length, placement, and density of (poroelastic) tentacles/oral arms. For our study in Section \ref{results:sec_length} we will vary the tentacle/oral arm length at $Re=150$. Finally in Section \ref{results:sec_density} we will vary the number, placement, and density of tentacles/oral arms. 

\begin{table}
\begin{center}
\begin{tabular}{| c | c | c | c |}
    \hline
    Parameter               & Variable    & Units        & Value \\ \hline
    Domain Size            & $[L_x,L_y]$  & m               &  $[5,12]$             \\ \hline
    Spatial Grid Size      & $dx=dy$      & m               &  $L_x/320=L_y/768$            \\ \hline
    Lagrangian Grid Size    & $ds$        & m               &  $dx/2$               \\ \hline
    Time Step Size          & $dt$        & s               &  $10^{-5}$   \\ \hline
    Total Simulation Time    & $T$        & \textit{pulses} &  $8$               \\ \hline
    Fluid Density            & $\rho$     & $kg/m^3$        &  $1000$               \\ \hline
    Fluid Dynamic Viscosity & $\mu$      & $kg/(ms)$       &  \textit{varied}      \\ \hline
    Bell Radius           & $a$        & m               &  $0.5$  (and \textit{varied})  \\ \hline
    Bell Diameter           & $D$ ($2a$)        & m        &  $1.0$  (and \textit{varied}) \\ \hline
    Bell Height           & $b$        & m               &  $0.75$   \\ \hline
    Contraction Frequency   & $f$       & $1/s$          &  $0.8$     \\ \hline
    Spring Stiffness   & $k_{spr}$ & $kg\cdot m/s^2$ &  $1\times10^{7}$  \\ \hline
    Beam Stiffness   & $k_{beam} $ & $kg\cdot m/s^2$ &  $2.5\times10^{5}$  \\ \hline
    Muscle Spring Stiffness    & $k_{muscle}$ & $kg\cdot m/s^2$ &  $1\times10^{5}$\\ \hline
    Poroelasticity Coefficient & $\alpha$     & $m^{-2}$        & \textit{varied} \\ \hline
    \end{tabular}
    \caption{Numerical parameters used in the two-dimensional simulations.}
    \label{table:num_param}
    \end{center}
\end{table}

For all studies, the computational width was kept constant as $L_x=5.$ Convergence studies were previously conducted in Miles et al. 2019 \cite{Miles:2019} that demonstrated low relative errors in forward swimming speeds for domain sizes from $L_x\in[3,8]$ for $Re=\{37.5,75,150,300\}$. In all cases, a similar trend was observed where narrower computational domains lead to slightly decreased forward swimming speeds while qualitative differences vortex formation were minimal. Additional grid resolution convergence studies were performed in Battista et al. 2019 \cite{Battista:2019} using the same computational model of a $2D$ jellyfish (but without tentacles/oral arms) that demonstrate appropriate error tolerances for the numerical parameters described in Table \ref{table:num_param}.

%
%

\subsection{Jellyfish Computational Model}
\label{sec:jelly_model}

\begin{figure}[H]
\centering
\includegraphics[width=0.75\textwidth]{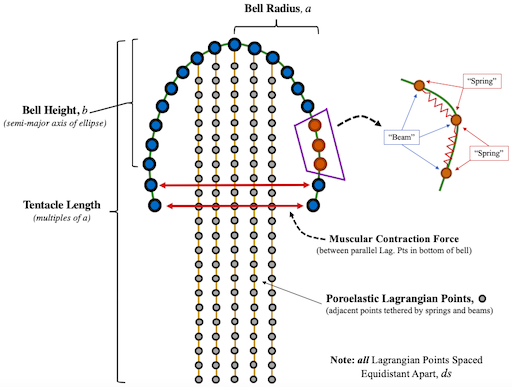}
\caption{Jellyfish model geometry composed of discrete points is a semi-elliptical configuration with tentacles/oral arms. The points along the bell are connected by virtual springs and virtual beams and the tentacles/oral arms are modeled as poroelastic structures, which include virtual springs and beams tethering adjacent points in the \textit{IB2d} software.}
\label{fig:Model_Geometry}
\end{figure}

The geometry is composed of a semi-ellipse with semi-major axis, $b=0.75$, and semi-minor axis, $a=0.5$, along with tentacles/oral hands that hang within the interior of the bell about its center, see Figure \ref{fig:Model_Geometry}. We refer to the bell height and radius as the semi-major axis, $b$, and semi-minor axis, $a$, respectively. As shown, it is composed of discrete Lagrangian points that are equally spaced a distance $ds$ apart. We note that this \textit{Lagrangian} mesh is twice as resolved as the background fluid grid, e.g., $ds=0.5dx$. Also, as the tentacle number, length, location, and density were varied through Sections \ref{results:sec_Re}-\ref{results:sec_density}, additional model geometry figures will be presented to distinguish between different geometrical setups when appropriate. 

Although we view the jellyfish as being immersed in the fluid, the jellyfish (Lagrangian grid) and fluid (Eulerian mesh) only communicate through integral equations with delta function kernels (see Eqs. (\ref{eq:force1})-(\ref{eq:force2}) in Appendix \ref{IB_Appendix} for more details.) In a nutshell, the Lagrangian mesh is allowed to move and deform. As an outside observer, we can witness this during the simulation. On the other hand, the Eulerian mesh is constrained upon a specific discrete rectangular lattice. On this mesh we are only measuring fluid related quantities, e.g., its velocity, pressure, or external forces upon it. One can envision the Eulerian mesh as though we have placed a number of measuring devices at the lattice points and can only track fluid quantities at those points; we are not tracking individual fluid blobs. 

The elegance of IB lies with how the Lagrangian grid and Eulerian mesh communicate to one another - through integral equations with delta function kernels (see Eqs. (\ref{eq:force1})-(\ref{eq:force2}) in Appendix \ref{IB_Appendix}). Simply put, these integrals say that the fluid points (on the rectangular) nearest the jellyfish (on the moving Lagrangian mesh) are influenced most by the jellyfish's movement (via a force), while the fluid grid points further away feel substantially less (Eq. (\ref{eq:force1})). A similar analogy can be made that says the fluid motion that influences the deformations/movement of the jellyfish the most are the fluid grid points nearest the jellyfish (Eq. (\ref{eq:force2})).

Successive points along the jellyfish bell are tethered together by \textit{virtual springs} and \textit{virtual (non-invariant) beams}, as well as along each tentacle/oral arm, in the \textit{IB2d} framework, as illustrated in Figure \ref{fig:Model_Geometry}. Virtual springs allow the geometrical configuration to either stretch or compress, while virtual beams allow bending between three successive points. When the geometry stretches/compresses or bends there is an elastic deformation force that arises from the configuration not being in its preferred energy state, e.g., its initial configuration. 

These deformations forces can be computed as below,
\begin{align}
\label{eq:spring} &\text{\footnotesize $\textbf{F}_{spr} = k_{spr} \left( 1 - \frac{R_L(t)}{||{\bf{X}}_{A}(t)-{\bf{X}}_{B}(t)||} \right) \cdot \left( \begin{array}{c} x_{A}(t) - x_{B}(t) \\ y_{A}(t) - y_{B}(t) \end{array} \right) $}\\
    \label{eq:beam} &\textbf{F}_{beam} = k_{beam} \frac{\partial^4}{\partial s^4} \Big( \textbf{X}_C(t) - \textbf{X}^{con}(t) \Big),
\end{align}
where $k_{spr}$ and $k_{beam}$ are the spring and beam stiffnesses, respectively, $R_L(t)$ are the springs resting lengths (set to $ds$, the distance between successive points), $\textbf{X}_{A}=\langle x_A,y_A\rangle$ and $\textbf{B}=\langle x_B,y_B\rangle$ are the Lagrangian nodes tethered by the spring. Note that $R_L(t)$ indicates that the resting lengths could be time dependent.  $\textbf{X}_C(t)$ is a Lagrangian point on the interior of the jellyfish bell and $\textbf{X}^{con}(t)$ is the corresponding initial (preferred) configuration of that particular Lagrangian point on the jellyfish bell. The spring stiffness are large to ensure minimal stretching or compression of the jellyfish bell itself, although via the beam formulation it is capable of bending and hence contracting. The $4^{th}$-order derivative discretization for the non-invariant beams is given in \cite{BattistaIB2d:2018}. Note that the beams are deemed non-invariant because the preferred configuration is non-invariant under rotations, so if the jellyfish were to turn, the model dynamics would undergo undesired motion due to these artifacts.

In addition to the tentacles/oral arms being modeled using springs and beams, they are also modeled as poroelastic structures. As the tentacles/oral arms deform, fluid is then allowed to slip through them, based upon the magnitude of their deformation. The poroelasticity is based upon the Brinkman flow model which is extended to include a slip velocity based on the deformation of a Lagrangian structure \cite{BattistaIB2d:2018}. The Brinkman force term is recognized as $\mu\alpha(\textbf{x}) \textbf{u}(\textbf{x},t)$, where $\alpha(\textbf{x})$ is the inverse of the hydraulic permeability. This term is added onto the right-hand side of the momentum equation in the Navier-Stokes equations (see Eq.(\ref{eq:NS1}) in Appendix \ref{IB_Appendix}) and is traditionally used to model flows through porous media \cite{Brinkman:1949,Nield:1992}. 

Once the deformation forces along the tentacle/oral arms are calculated, they are equated to the Brinkman force term, which is then incorporated into a slip velocity, e.g.,

\begin{align}
    \label{eq:brink1} \textbf{f}_{brink} &= -\textbf{f}_{elastic} \\
    \label{eq:brink2} \textbf{U}_b(\textbf{x},t) &= \textbf{u}(\textbf{x},t) + \frac{\textbf{f}_{elastic}}{\alpha(\textbf{x})\mu}.
\end{align}

In our model $\alpha(\textit{x})=\alpha$ is spatially independent and was chosen as $\alpha=500,000, 10,000$ and $25,000$ for all simulations in Sections \ref{results:sec_Re}, \ref{results:sec_length}, and \ref{results:sec_density}, respectively. Note that varying the poroelastic coefficient, $alpha$, did not significantly affect forward swimming speeds (see Appendix \ref{app:varying_alpha}). Moreover, note that our model does not distinguish between tentacles or oral arms structurally and will from now on refer to them as tentacles/oral arms.

Lastly to mimic the subumbrellar and coronal muscles that induce bell contractions, virtual springs. These springs dynamically change their resting lengths in a sinusoidal fashion, see Eq.(\ref{eq:RL_muscle}). Rather than tether neighboring points with virtual springs to model the muscles, we tether points across the jellyfish bell using all Lagrangian points that are below the top hemi-ellipse. The deformation force equation does not change from Eq.(\ref{eq:spring}) besides using a different spring stiffness, which we call $k_{muscle}$, and time-dependent spring resting lengths. The equation that governs the time-dependent resting lengths, $R_L(t)$, is given as follows 
\begin{equation}
    \label{eq:RL_muscle} RL(t) = 2a\left( 1 - \left| \sin(\pi t f) \right| \right).
\end{equation}

\begin{figure}[H]
\centering
\includegraphics[width=0.99\textwidth]{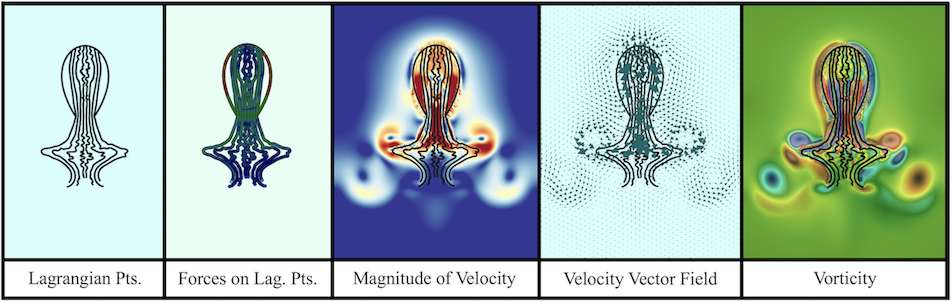}
\caption{A snapshot of a jellyfish simulation with $8$ tentacles/oral arms swimming at $Re=150$ during its $5^{th}$ contraction cycle, illustrating some of the simulation data obtained at each time-step, e.g., positions of Lagrangian points as well as forces on them, magnitude of velocity, the velocity vector field, and vorticity. Note other data not visualized is the fluid pressure and Lagrangian forces spread from the jellyfish onto the Eulerian (fluid) grid.}
\label{fig:example_sim_data}
\end{figure}

Upon running all simulations, we stored the following data in increments of 4\% of each contraction cycle:
\begin{enumerate}
    \item Position of Lagrangian Points
    \item Horizontal/Vertical/Normal/Tangential Forces on each Lagrangian Point
    \item Fluid Velocity 
    \item Fluid Vorticity
    \item Fluid Pressure
    \item Forces spread onto the Fluid (Eulerian) grid from the Jellyfish (Lagrangian) mesh
\end{enumerate}
We then used the open-source software called VisIt \cite{HPV:VisIt}, created and maintained by Lawrence Livermore National Laboratory for visualization, see Figure \ref{fig:example_sim_data}, and the data analysis package software within \textit{IB2d} \cite{BattistaIB2d:2017} for data analysis. Figure \ref{fig:example_sim_data} provides a visualization of some of the data produced from a single jellyfish locomotion at a single snapshot in time during its $5^{th}$ contraction cycle, for a jellyfish with $8$ tentacles (see Section \ref{results:sec_Re}) at a Reynolds Number of $150$.

We stored $25$ time points per contraction cycle in each jellyfish simulation and temporally-averaged data over the 7th and 8th (last two) contraction cycles to compute average swimming speed for a particular simulation, $U_{avg}$. Note that forward swimming speed steadied out by $6$ contraction cycles (see Figure \ref{fig:Re_DistanceVelocity} in Section \ref{results:sec_Re}). Moreover we the computed the Strouhal Number, $St$, 

\begin{equation}
    \label{eq:Str} St = \frac{fD}{U_{avg}}
\end{equation}
where $f$ is the contraction frequency, $D$ is the maximum bell diameter during a contraction cycle, and $U_{avg}$ is the temporally-averaged forward swimming speed. By taking the inverse of $St$ we can get a normalized forward swimming speed based on driving frequency. Previous studies on animal locomotion have hypothesized that propulsive efficiency is high in a narrow band of $St$, peaking within the interval $0.2 < St < 0.4$ \cite{Taylor:2003}. We are particularly interested in exploring the any relationships between $Re$, $St$, and tentacle/oral arm number density and placement. Furthermore, we will also compute the cost of transport (COT) and investigate its relationship to the aforementioned parameters. The cost of transport has been used as a measure of aquatic locomotion efficiency and is a measure of energy (or power) spent per unit distance traveled \cite{Schmidt:1972,Bale:2014}. We computed both a power-based COT and energy-based (work) COT, e.g.,

\begin{align}
    \label{eq:COTwork} COT_{work} &= \frac{1}{N} \frac{1}{d_S} \displaystyle\sum_{j=1}^N |F_j||d_j| \\
    \label{eq:COTPower} COT_{power} &= \frac{1}{N} \frac{1}{d_S} \displaystyle\sum_{j=1}^N |F_j||U_j|,
\end{align}
where $F_j$ is the applied contraction force at the $j^{th}$ time step by the jellyfish, $U_{r_{j}}$ is is the bell contraction velocity at the $j^{th}$ time step, $d_j$ is the lateral distance the bell contracted (or expanded) during the $j^{th}$ time step, $N$ is the total number of time steps considered, and $d_S$ is the distance swam by the jellyfish traveled during during this period of time across all time steps considered.

Lastly, we computed the finite-time Lyapunov exponents (FTLE) which is used as a metric for instantaneous Lagrangian Coherent Structures (LCS) \cite{Haller:2000,Shadden:2005,Haller:2015}. In a nutshell, LCSs provide a systematic way to untangle the intricate, complex, and hidden dynamics of the system in a way that can be clearly visualized and interpreted. Within fluid flows, LCSs help reveal particle transport patterns that are of particular importance in biology. They can be used to understand various processes for jellyfish, including feeding and prey-capture \cite{Peng:2009,Sapsis:2011} and locomotion \cite{Franco:2007,Zhang:2008,Wilson:2009,Lipinski:2009,Haller:2011,Taheri:2018,Miles:2019}. FTLEs were computed using VisIt \cite{HPV:VisIt}, where trajectories were computed using instantaneous snapshots of the 2D fluid velocity vector field on the entire computational domain using a forward/backward Dormand-Prince (Runge-Kutta) time-integrator with a relative and absolute tolerance of $10^{-4}$ and $10^{-5}$, respectively, and a maximum advection time of $0.02s$ (1.6\% of a contraction cycle) with a maximum number of steps of 250. They were visualized using a colormap corresponding to the FTLE value on the background grid as well as FTLE contours.

\section{Results}

\begin{figure}[H]
\centering
\includegraphics[width=0.7\textwidth]{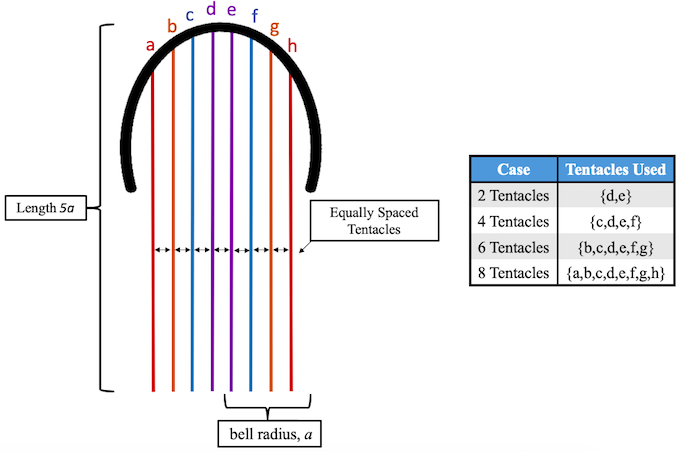}
\caption{Geometric model considered in Section \ref{results:sec_Re} to determine how the presence of tentacles/oral arms affects forward swimming speed. This same geometry is used in Section \ref{results:sec_length} but with different tentacle/oral arms lengths, given in multiples of the bell radius, $a$.}
\label{fig:Re_Geo}
\end{figure}

Using an extension of an idealized jellyfish model by Hoover et al. 2015 \cite{Hoover:2015}, and later modified by Miles et al. 2019 \cite{Miles:2019}, we incorporated poroelastsic structures that mimic tentacles/oral arms and investigated how such complex tentacle/oral arm structure affects forward swimming performance, the cost of transport (COT), and differences in Lagrangian Coherent Structures (LCS). We first sought out to explore the relationship between $Re$ and the number of symmetric tentacles/oral arms. Next we chose four specific $Re$ (37.5, 75, 150, and 300), and investigated effects of tentacle/oral arm length using the same symmetric tentacle/oral arm geometry, but with differing lengths of the tentacles/oral arms, see Figure \ref{fig:Re_Geo}.

We then explored the how the placement and density of tentacles/oral arms affects forward swimming performance and cost of transport across $Re=\{37.5,75,150,300\}$. Thus we performed three sub-studies to investigate the following questions:
\begin{enumerate}
    \item Is it only the outer placed tentacles that affect its swimming?
    \item How does tentacle density affect its swimming? 
    \item If the placement of interior tentacles is varied, will it affect its swimming?
\end{enumerate}

For each study we focused on quantifying differences in forward swimming speed (and Strouhal Number, $St$) and COT. We also investigated differences in Lagrangian Coherent Structures (LCS) formation using finite-time Lyanunov Exponents (FTLE) analysis to decipher regions of mixing and where fluid is being pulled/pushed in Sections \ref{results:sec_Re}-\ref{results:sec_length} to explore dynamical differences due to including tentacles/oral arms.

%
%

\subsection{Results: Varying $Re$ and Number of Tentacles/Oral Arms}
\label{results:sec_Re}

\begin{figure}[H]
\centering
\includegraphics[width=0.99\textwidth]{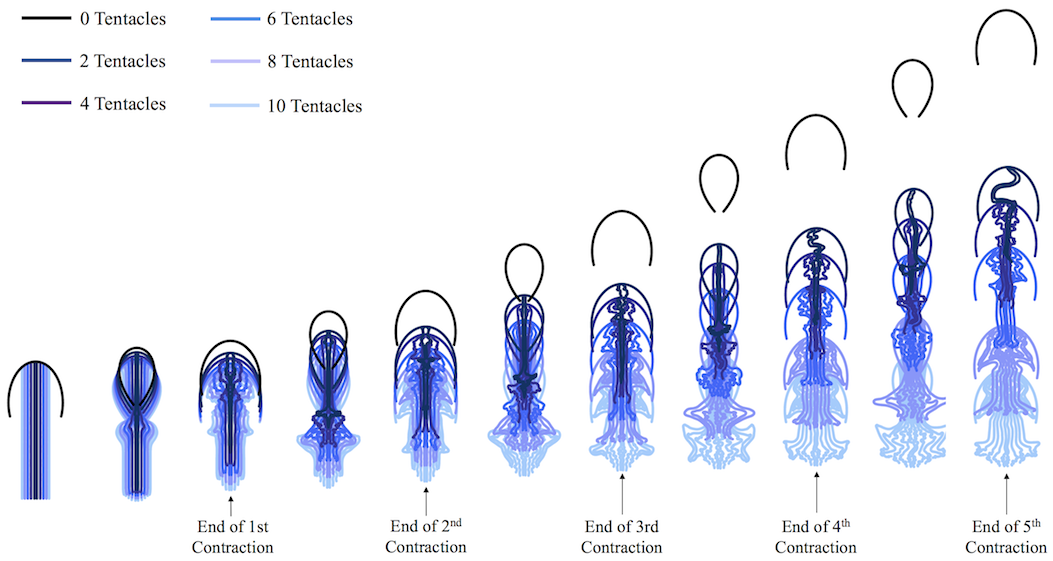}
\caption{Visualization comparing jellyfish swimming for a variety of different number of symmetric tentacles/oral arms, for $Re=150$ with a contraction frequency of $f=0.8\ Hz$. As the number of tentacles increases, forward swimming progress is more limited.}
\label{fig:TentNum_Swim_Compare}
\end{figure}

Previous studies of jellyfish locomotion have considered forward swimming performance over a range of $Re$ \cite{Hershlag:2011,Yuan:2014,Miles:2019}; however, all have only modeled the jellyfish bell without any additional complex morphology, such as tentacles/oral arms. Here we consider the jellyfish geometries presented in Figure \ref{fig:Re_Geo} to investigate how tentacles/oral arms may affect forward swimming speeds.  To explore further, we placed the \textit{in silico} jellyfish in increasingly more viscous fluids while holding all other parameters constant to effectively decrease the $Re$. We observed that forward swimming performance decreases, even with the addition of tentacles/oral arms. 

\begin{figure}[H]
\centering
\includegraphics[width=0.975\textwidth]{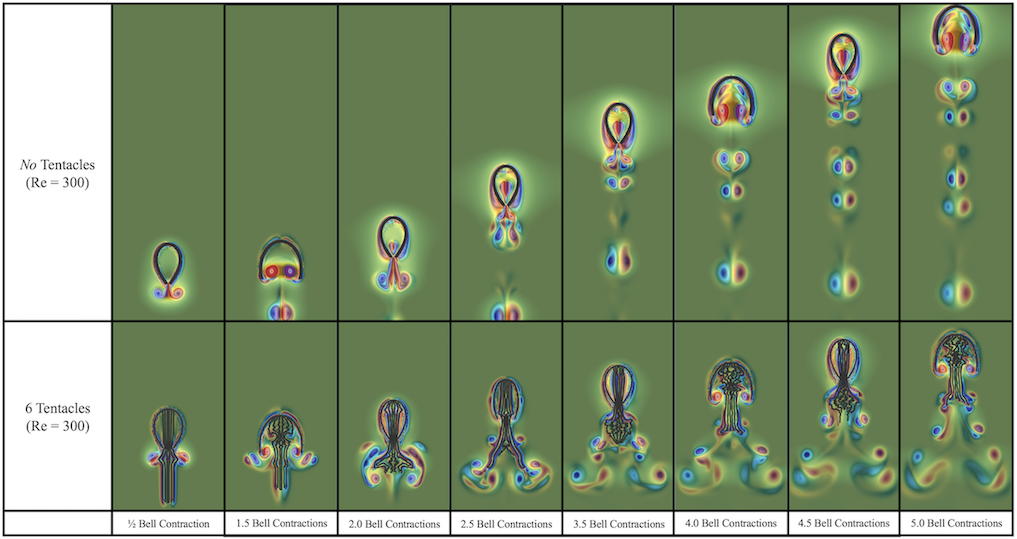}
\caption{Visualization comparing a jellyfish with no tentacles/oral arms to the case with 6 tentacles/oral arms (3 symmetric per side) at $Re=150$. The colormap represents vorticity.}
\label{fig:Compare_0_6_tents}
\end{figure}

\begin{figure}[H]
\centering
\includegraphics[width=0.975\textwidth]{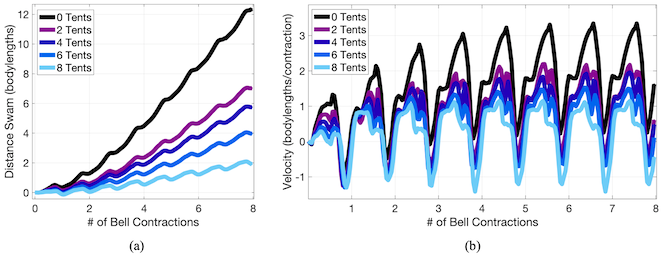}
\caption{Plots detailing (a) distance swam and (b) velocity over 8 bell contraction periods at $Re=150$ for differing numbers of symmetric tentacles.}
\label{fig:Re_DistanceVelocity}
\end{figure}

This phenomenon is visualized in Figures \ref{fig:TentNum_Swim_Compare}, \ref{fig:Compare_0_6_tents}, and \ref{fig:Re_DistanceVelocity}, where the first illustrates that at $Re=150$  including more tentacles/oral arms decreases forward swimming distance, the second shows dynamical differences in the vortex wake at for a case with 0 and 6 tentacles/oral arms at $Re=150$, and the last visualizes the distance swam and swimming velocity for the first 5 bell contractions performed for differing numbers of tentacles at $Re=150$. This data suggests that jellyfish with lesser numbers of tentacles/oral arms have higher forward swimming speeds and also that the presence of tentacles/oral arms suppresses vortex formation. Note that in this case here including more tentacles/oral arms means placing more further out within the bell, see Figure \ref{fig:Re_Geo}.

\begin{figure}[H]
\centering
\includegraphics[width=0.975\textwidth]{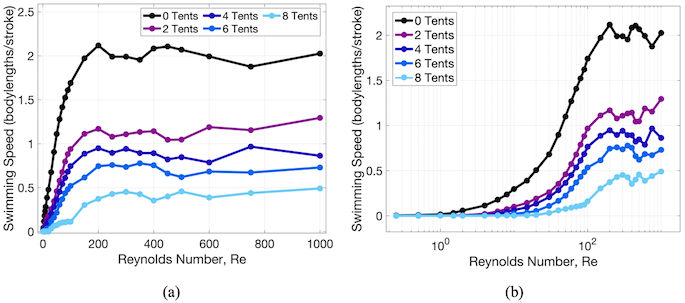}
\caption{Illustrating average forward swimming speed against Reynolds Number, $Re$, for different number of symmetric tentacles/oral arms. Swimming speed is measured in non-dimensional units (bodylengths/contraction) in normal form (a) and logarithmic form (b). It is clear that the addition of tentacles/oral arms decreases forward swimming performance.}
\label{fig:Re_Speed}
\end{figure}

Figure \ref{fig:Re_Speed} explores these relationships further by quantifying forward swimming speeds across different $Re$, ranging from $0.25$ to $1000$, for varying numbers of symmetrically placed tentacles. Both Figure \ref{fig:Re_Speed}$a$ and $b$ presents the same data, but $b$ uses a semi-logarithmic $Re$ axis. In agreement with previous models, forward swimming is negligible for $Re\lesssim1$ and significant forward swimming begins around $Re\gtrsim10$, when inertial effects become greater than viscous dampening. See Figure \ref{fig:ReSpeedLogLog} in Appendix \ref{app:varying_Re} to more clearly observe the speeds at lower $Re$.

Moreover, as viscosity decreases (and $Re$ increases) forward swimming performance, e.g., swimming speed, increases for $10\lesssim Re\lesssim300$, similar to \cite{Miles:2019}. Note that the lower end of this range various slighty for each case of differing tentacle number. For $Re\gtrsim150$, forward swimming speed steadies out regardless of the number of symmetrically placed poroelastic tentacles/oral arms, see Figure \ref{fig:Re_Speed}. The addition of tentacles/oral arms monotonically decrease forward swimming performance. At higher $Re$ we see similar behavior to the case with no tentacles/oral arms, in which swimming speed asymptotically steadies out. Upon comparing the case with $0$ tentacles/oral arms to $8$ tentacles (4 symmetrically placed per side), the case with $0$ is $\sim400\%$ faster than the $8$ tentacle case!

Next we investigated the relationship between $Re$, tentacle/oral arm number, and Strouhal Number, $St$. By reporting the forward swimming speed in non-dimensional terms, given in bodylengths per bell contraction, it is the inverse of $St$. The data illustrates that the inclusion of poroelastic tentacle/oral arms increases $St$ (since speed is slower), where for a given $Re$ more tentacles/oral arms increases $St$. The only case here that falls within the biological regime of $0.2<St<0.4$ \cite{Taylor:2003} is the case with no tentacles/oral arms when $Re\gtrsim 50$. Furthermore, in terms of scaling relations, even with the addition or tentacles/oral arms, the $St$ is a monotonically decreasing function of $Re$ before steadying out around $Re\gtrsim150$, suggesting that increasing $Re$ maximizes propulsive efficiency \cite{Nudds:2014} agreeing with previous jellyfish studies of \cite{Miles:2019}.

\begin{figure}[H]
\centering
\includegraphics[width=0.5\textwidth]{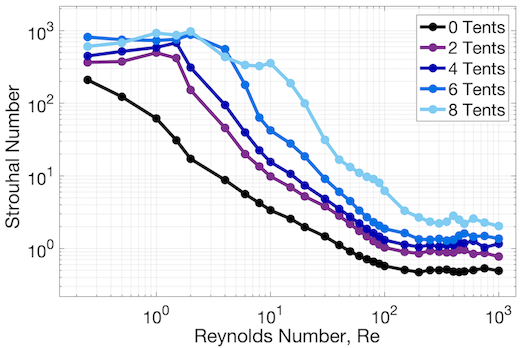}
\caption{Plot depicting the relationship between Strouhal Number, $St$, and Reynolds Number, $Re$, for different numbers of symmetric tentacles/oral arms. $St$ is the inverse of non-dimensional swimming speed.}
\label{fig:Re_Strouhal}
\end{figure}

\begin{figure}[H]
\centering
\includegraphics[width=0.975\textwidth]{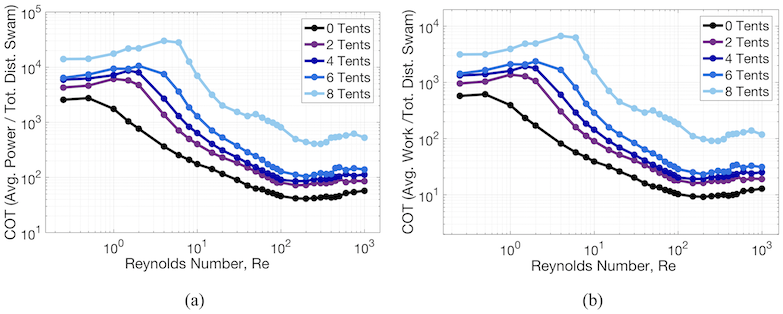}
\caption{Illustrating the relationship between cost of transport (COT) and Reynolds Number, $Re$, for different numbers of symmetric tentacles/oral arms, when COT is computed using (a) average power and (b) average work.}
\label{fig:Re_COT}
\end{figure}

Experimental studies of jellyfish have concluded that the cost of transport (COT) for jellyfish is much lower than other metazoans \cite{Gemmell:2013}. They described this phenomenon by suggesting that jellyfish use passive energy recapture to decrease COT. It was later studied computationally by Hoover et al. 2019 \cite{Hoover:2019}, confirming this as the reason for lower COT in a model. We hypothesized that the scaling relationship between cost of transport, number of tentacles/oral arms, and $Re$ would be similar to those shown in Miles et al. 2019 \cite{Miles:2019}, where higher $Re$ lead to decreases COT for a variety of contraction frequencies; however, where for a specified number of tentacles/oral arms the COT would increase due to slower forward swimming speeds, where the kinematics of contraction remain uniform between all tentacle/oral arm cases. Figure \ref{fig:Re_COT} highlights the COT data for both an average work-based and average power-based COT over multiple orders of magnitude of $Re$ and varying numbers of tentacles/oral arms. Generally, COT decreases as $Re$ increases. For a given $Re$, the highest COT corresponds to the case with the most tentacles/oral arms (similar to Figure \ref{fig:Re_Strouhal}). Note that for a particular jellyfish morphology (with a specified number of tentacles/oral arms), optimal swimming performance occurs near $Re\gtrsim100$, where COT appears to be minimal and forward swimming speed is maximized.

\begin{figure}[H]
\centering
\includegraphics[width=0.975\textwidth]{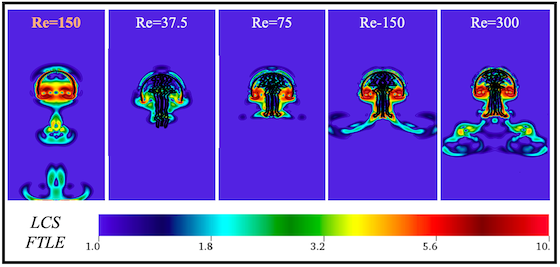}
\caption{Visualization comparing Lagrangian Coherent Structures (LCS) using finite-time Lyanpunov exponents (FTLE) for the case with 6 total tentacles/oral arms (3 symmetrically placed per side) and $Re=\{37.5,75,150,300\}$ at the beginning of the $4^{th}$ contraction cycle. Note that the case of $Re=150$ with no tentacles is given to provide a comparison.}
\label{fig:LCS_Re_1T}
\end{figure}

We then performed Lagrangian Coherent Structure (LCS) analysis by computing the finite-time Lyapunov exponent (FTLE) \cite{Haller:2000}. Small values of the FTLE highlight regions where flow is attractive while large values indicate areas in which the flow is repelling. For jellyfish, LCSs can be used to highlight the regions of fluid in which the jellyfish is pulling towards or pushing away from its bell during contraction and expansion, respectively \cite{Wilson:2009}. Figure \ref{fig:LCS_Re_1T} compares the FTLE LCS analysis at the start of the $4^{th}$ contraction cycle between  $Re=\{37.5,75,150,300\}$ for the case of $6$ total tentacles/oral arms (3 symmetrically placed on each side of the bell). To see a comparison over the entire contraction cycle, see Figure \ref{fig:LCS_Re} in Appendix \ref{app:varying_Re}. During bell contraction, high FTLE values are seen near the tips of the jellyfish bell, indicating regions of high fluid mixing in all cases. For cases with $Re\gtrsim 75$, those regions on high fluid mixing are expelled downward by the time the contraction phase ends; however, in comparison to the case with no tentacles (see \cite{Miles:2019}), there is less mixing in the vortex wake due to suppressed vortex formation. Moreover, there appears to be further horizontal mixing in the wake in comparison. In general as $Re$ increases, the size of the regions with high FTLE also increase in the vortex wake.

\begin{figure}[H]
\centering
\includegraphics[width=0.975\textwidth]{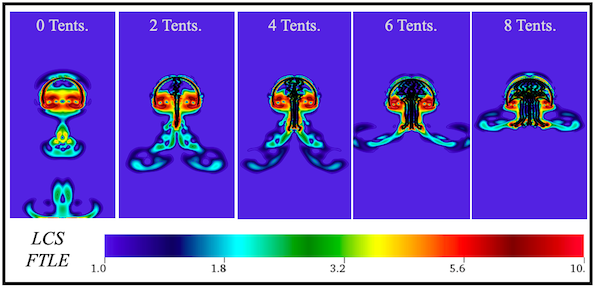}
\caption{Visualization comparing Lagrangian Coherent Structures (LCS) using finite-time Lyanpunov exponents (FTLE) for cases with either $0,1,2,3$ or $4$ symmetrically-placed tentacles/oral arms per side for $Re=150$ at the start of the $4^{th}$ contraction cycle.}
\label{fig:LCS_TentNumSweep_1T}
\end{figure}

In each case, the low FTLE values above the bell suggest that fluid is being pulled downwards towards the end of the bell during contraction, rather than the jellyfish horizontally pulling in fluid, even with the addition of tentacles. Within the bell, similar to \cite{Miles:2019}, fluid is pulled inwards and towards the top during contraction and expansion; however, there is much less pronounced mixing within the center of the bell due to the presence of tentacles/oral arms. The high FTLE values near the bottom of the bell that suggest highly-attractive flow regions could be vital for allowing jellyfish to expel wastes from its mouth and out of their bells during a contraction cycle. Furthermore, more tentacles/oral arms appear to decrease the amount of mixing in the vortex wake, see Figure \ref{fig:LCS_TentNumSweep_1T}. There are still higher levels of mixing near the bottom of the bell with the addition of more tentacles/oral arms. To observe the effect of tentacle number on fluid mixing see Figure \ref{fig:LCS_TentNumSweep} in Appendix \ref{app:varying_Re}.

For this particular jellyfish geometry, these results suggest that the jellyfish would ideally want its prey to be in front (a top) of its bell so that during a contraction cycle, the prey would be pulled downwards towards its bell and into its tentacles. These results could be specific to this jellyfish bell geometry. It is possible that changes in bell diameter or contraction kinematics could lead to differences in fluid mixing patterns, suggesting prey are captured more horizontally rather than vertically, like in the case of \textit{Cassiopea} \cite{Santhanakrishnan:2012,Hamlet:2012}.

Next we explored how the length of tentacles/oral arms affects forward swimming.


%
%
%
%

\subsection{Results: Varying the Length of Tentacles/Oral Arms}
\label{results:sec_length}

While in Section \ref{results:sec_Re} we observed that the number of tentacles significantly affects locomotive processes at various fluid scales ($Re$), we did not vary the length of the tentacles/oral arms. In this section we will fix the Reynolds Number at $Re=150$, which is approximately the $Re$ of \textit{Sarsia} and vary the number of tentacles/oral arms and their respective length. Note that we use the model geometry given in Figure \ref{fig:Re_Geo} but with varying lengths of the tentacles/oral arms. As this is a first study of tentacles/oral arms we will still use uniform tentacle/oral arm length in each respective case. Throughout this section the length of a tentacle is given in multiples of the bell radius, $a$.

\begin{figure}[H]
\centering
\includegraphics[width=0.5\textwidth]{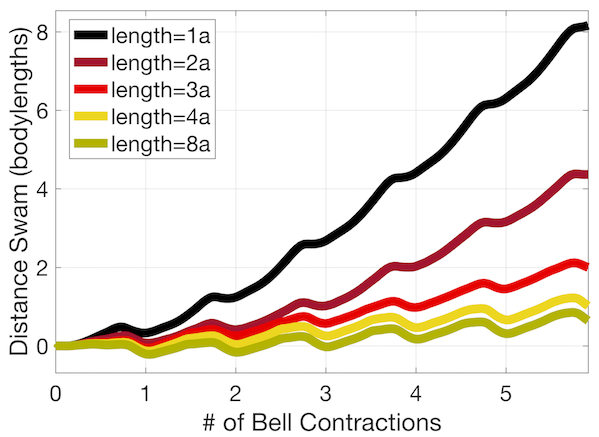}
\caption{Plot detailing distance swam against bell contractions performed for differing tentacle/oral arm lengths at $Re=150$. Tentacle/oral arm length is given in multiples of the bell radius, $a$.}
\label{fig:Len_Distance}
\end{figure}

First we observed that longer tentacle/oral arms leads to decreased forward distance swam, see Figure \ref{fig:Len_Distance}. Figure \ref{fig:Len_Distance} gives the distance swam against number of bell contractions for the case of $8$ tentacles/oral arms ($4$ per side). As tentacle/oral arm length increases, the jellyfish does not travel as far, thus it appears if they are short enough, they may not significantly change swimming performance from the case in which there are none. As length increases, forward swimming gets decreasingly less pronounced; however, when the lengths get long enough ($\sim 4a$), it appears elongating tentacles/oral arms further will not significantly decrease forward swimming performance, as the data appears to asymptotically steady out (see Figure \ref{fig:Len_Distance} and compare with the $8a$ case). 

\begin{figure}[H]
\centering
\includegraphics[width=0.975\textwidth]{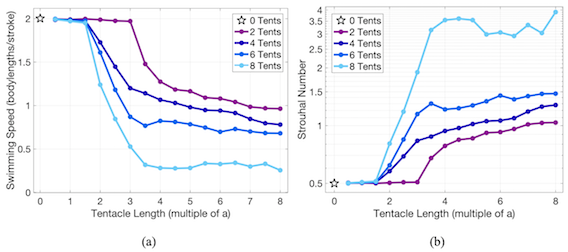}
\caption{Illustrating average forward swimming speed for different lengths of symmetrically placed tentacles/oral arms at $Re=150$. Swimming speed is measured in non-dimensional units ($bodylengths/contraction$) and tentacle length is measured in multiples of the bell radius, $a$. As tentacle/oral length increases, forward swimming speed decreases, until it appears to steady out.}
\label{fig:Len_Speed}
\end{figure}

This last idea is further solidified when quantifying forward swimming speeds, see Figure \ref{fig:Len_Speed}a. Figure \ref{fig:Len_Speed}a provides the forward swimming speed for multiple tentacle/oral arm lengths (given in multiples of the bell radius, $a$) for cases involving different numbers of tentacle/oral arms. If length is short enough, even in cases of $8$ total tentacles/oral arms ($4$ per side), swimming speeds are not significantly different from the case with no tentacles/oral arms. In fact, the data appears to converge towards the case with none as length gets shorter. 

As length increases, the number of equally-spaced and symmetrically-placed tentacles/oral arms begins to matter. That is, the swimming speeds between all cases begins to diverge around $\sim 1.5a$. While swimming speed dramatically decreases in the case of $8$ tentacles/oral arms, for the case with $2$ tentacles/oral arms the forward swimming speed is not too different from the case with no tentacles/oral arms. This may be due due to the $8$ tentacles/oral arms taking up more space within the bell, which seems to suppress vortex formation and thus reduces the size of the vortex ring expelled during expansion. Thus, each number of tentacles/oral arms the forward swimming speed decreases at different rates as a function of the tentacle/oral arm length. As length increases, eventually swimming speeds appear to steady out, and elongating them further will not significantly decrease swimming speed. 

\begin{figure}[H]
\centering
\includegraphics[width=0.975\textwidth]{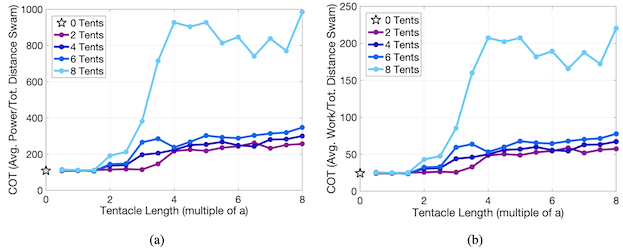}
\caption{Illustrating the relationship between cost of transport (COT) and tentacle/oral arm length for different numbers of symmetric tentacles/oral arms at $Re=150$, when COT is computed using (a) average power and (b) average work. Tentacle/oral arm length is given in multiples of the bell radius, $a$.}
\label{fig:Len_COT}
\end{figure}

Figure \ref{fig:Len_Speed}b gives the Strouhal Number, $St$, vs tentacle/oral arm length (quantified in multiples of the bell radius, $a$). For shorter tentacles/oral arms the $St$ converges to approximately 0.5, near the biological range of $0.2<St<0.4$; however for lengths greater than $\sim 2a$, $St$ grows of this range in all cases except case with $2$ tentacles/oral arms ($1$ per side), which grows after $\sim3a$. As length increases the $St$ continues to increases in each case, but appears to asymptotically taper out. A similar trend is seen when investigating the cost of transport ($COT$), see Figure \ref{fig:Len_COT}. Shorter tentacles/oral arms have lower $COT$, while longer have higher. As the number of the tentacles/oral arms increases, so does the $COT$. 


In each case of different numbers of tentacles/oral arms, there appears to be 3 different regimes of tentacle/oral arm length. (1) If the length is short enough ($\lesssim 1.5$ bell radii), forward swimming speeds are not significantly different than the case of no tentacles/oral arms. (2) Within a certain length range ($\sim1-2$ bell diameters), swimming speeds significantly drop off (3) For long enough tentacles/oral arms, swimming speed begins to steady out ($\gtrsim 2$ bell diameters). Coupling these ideas with the $COT$ data, lesser numbers of and shorter tentacles/oral arms suggest that these jellyfish may be better and more efficient active hunters (predators) than its jellyfish counterparts with more and longer tentacles/oral arms. One example of this could be box jellyfish (\textit{Carukia barnesi}) rather than a jellyfish with longer tentacles, such as the lion's mane (\textit{Cyanea capillata}) \cite{Linnaeus:1758,Kinsey:1988,Powell:2018}.

\begin{figure}[H]
\centering
\includegraphics[width=0.925\textwidth]{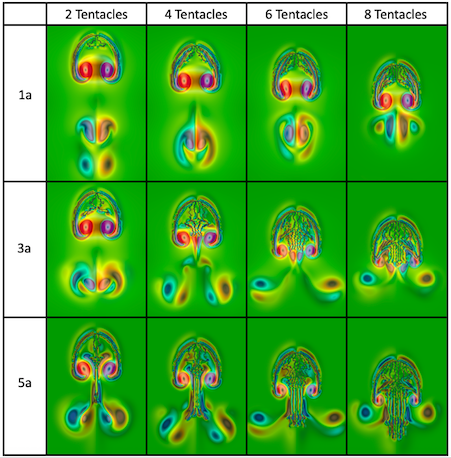}
\caption{Visualization of jellyfish position and a colormap of vorticity at the end of the $5^{th}$ contraction cycle for each case of differing number of tentacles/oral arms of specified length, at $Re=150$. Note that length is given in multiples of the bell radius, $a$.}
\label{fig:Len_Vorticity}
\end{figure}

The decreases in forward swimming speeds may be attributed to suppressed vortex formation and ring dynamics, as suggested previously in Figure \ref{fig:Compare_0_6_tents} and now Figure \ref{fig:Len_Vorticity}. Figure \ref{fig:Len_Vorticity} provides a colormap of vorticity at the end of the $5^{th}$ contraction cycle for cases of varying tentacles/oral arm number and lengths. For the case of having lengths $1a$, it is clear that the addition of more equally-spaced and symmetrically-placed tentacles/oral arms decreases the size of the downstream vortex wake. In particular, for cases of more tentacles/oral arms, vortex rings dissipate closer to the jellyfish than in the cases with lesser numbers. As the tentacle/oral arm length increases, the topology of the vortex wake is significantly altered. Rather than vortex rings being pushed directly downward, they begin to expel more horizontally, more laterally. Higher number and longer tentacles/oral arms appear to act, in essence, as an inelastic structure bouncing vortices back off of them, not allowing them to move directly downward.

\begin{figure}[H]
\centering
\includegraphics[width=0.99\textwidth]{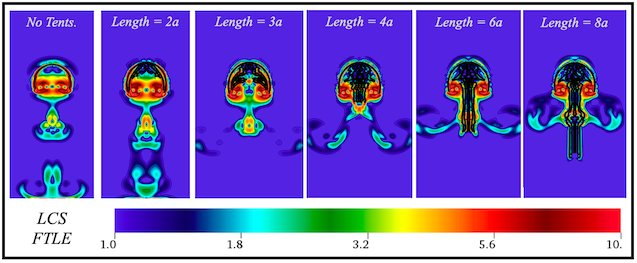}
\caption{Visualization comparing Lagrangian Coherent Structures (LCS) using finite-time Lyanpunov exponents (FTLE) for the case with 6 total tentacles/oral arms (3 symmetrically placed per side) of varying lengths (in multiples of the bell radius, $a$, at the start of the $4^{th}$ contraction cycle.}
\label{fig:LCS_Len_1T}
\end{figure}

These ideas are further explored while performing LCS analysis (via computing the finite time Lyapnuov exponents (FTLE)). We explored how regions of fluid mixing varied due to variations of tentacle/oral arm length, see Figure \ref{fig:LCS_Len_1T}. Figure \ref{fig:LCS_Len_1T} gives the FTLE as a colormap at the start of the $4^{th}$ contraction cycle for variety of tentacle/oral arm lengths. If lengths are short enough $(\lesssim 2a)$ there is still significant mixing downstream in a vertically longer vortex wake, as suggested by Figure \ref{fig:Len_Vorticity}. However, as lengths increases there is more horizontal mixing near the bottom of the bell, than downstream. Interestingly, in cases for lengths $\sim8a$ the ends of the tentacles/oral arms do not experience much mixing at all. This further suggests when there long enough and enough tentacles/oral arms in number, they act as an inelastic pseudo-wall bouncing vortices back off of them, which in turn causes them to move laterally, rather than in the opposite direction of swimming motion. The increase of lateral mixing within the tentacle/oral arm region would help the jellyfish capture prey in its tentacles for feeding.  To observe fluid mixing over the $4^{th}$ to $5^{th}$ contraction cycle see Figure \ref{fig:LCS_Len} in Appendix \ref{app:varying_length}.

Underlying the above results and discussion was the assumption that all tentacles/oral arms in Sections \ref{results:sec_Re} and \ref{results:sec_length} were equally-spaced and symmetrically-placed on each side of the jellyfish bell. That is, if we considered a jellyfish with $8$ tentacles/oral arms ($4$ per side), the geometry was exactly the same as the jellyfish with $6$, but with the addition of one more tentacle/oral arm per side, equally-spaced from the previously outermost. We will now relax this, and consider how changes in tentacle/oral density affects the swimming performance metrics studied above.

%
%
%
%

\subsection{Results: Tentacle/Oral Arm Placement and Density}
\label{results:sec_density}

In Sections \ref{results:sec_Re} and \ref{results:sec_length} we explored an idealized jellyfish model where all tentacles/oral arms were equally spaced apart from one another. For example, if a jellyfish had $6$ tentacles, four of its tentacles/oral arms would be placed exactly where the $4$-Tentacle/oral arm jellyfish had theirs placed, and then in addition to those four, it would have two additional tentacles/oral arms equally-spaced outside of those $4$, one per side. Here we will relax those assumptions and consider different placements and densities of the tentacles/oral arms inside a particular tentacle/oral arm containing region to investigate how placement and density affect potential forward swimming performance. 

The density and locations of tentacles/oral arms are extremely diverse among different species of jellyfish, recall Figure \ref{fig:SpeciesDiversity}. This is a preliminary study investigating how density and placement may play a role in differing swimming behavior for an idealized jellyfish model. In particular, we will address the following three questions to probe the surface of how different placements/densities of tentacles/oral arms may affect forward locomotion:
\begin{enumerate}
    \item \textit{Is the placement of the outermost tentacles/oral arms what affects swimming performance?} - We will hold the location of the outermost tentacles/oral arms constant and change the location of the inner tentacles/oral arms, see Figure \ref{fig:batch1_geo}. 
    \item \textit{How does density of tentacles affect swimming performance?} - We will again hold the location of the outermost tentacles/oral arms constant, and vary the amount of other tentacles inside that region; however, in contrast to the question above, the spacing between tentacles/oral arms will change as more or less tentacles/oral arms are considered within that region, see Figure \ref{fig:batch3_geo}.
    \item \textit{How does stacking tentacle/oral arms towards the outermost ones affect swimming performance?} - We will hold the location of the outermost tentacles constant and place more tentacles towards the outermost tentacles/oral arms and observe how swimming performance is affected. In addition we will explore if there are clusters of tentacles/oral arms and how they may affect forward swimming performance, see Figure \ref{fig:batch2_geo}. 
\end{enumerate}

%
%

\subsubsection{Is the placement of the outermost tentacles/oral arms what affects swimming performance?}
\label{results:sec_batch1}

\begin{figure}[H]
\centering
\includegraphics[width=0.6\textwidth]{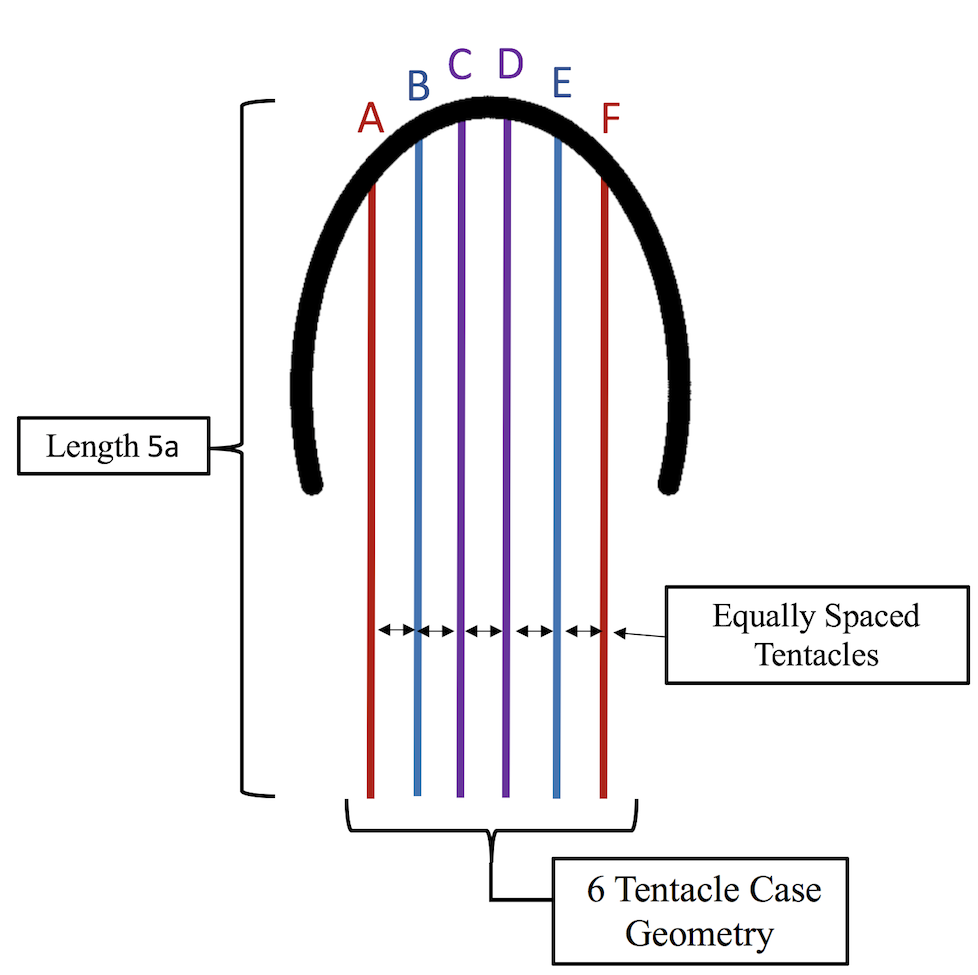}
\caption{Geometric setup for all cases considered in Section \ref{results:sec_batch1} to determine if the placement of the outermost tentacles/oral arms dictates forward swimming speed.}
\label{fig:batch1_geo}
\end{figure}

For this study, we will include the tentacles/oral arms labeled $A$ and $F$ in our jellyfish model across all simulations, see Figure \ref{fig:batch1_geo}. We will then consider four cases: (1) $ABCDEF$ (2) $ABEF$ (3) $ACDF$ and (4) $AF$. These cases vary in the number of total tentacles/oral arms, either 2, 4 or 6, as well as placement of inner tentacles/oral arms for $Re=37.5, 75, 150$ and $300$. Note that the case $ABCDEF$ is the same case as in the $6$ tentacle case in Section \ref{results:sec_Re}. 

\begin{figure}[H]
\centering
\includegraphics[width=0.975\textwidth]{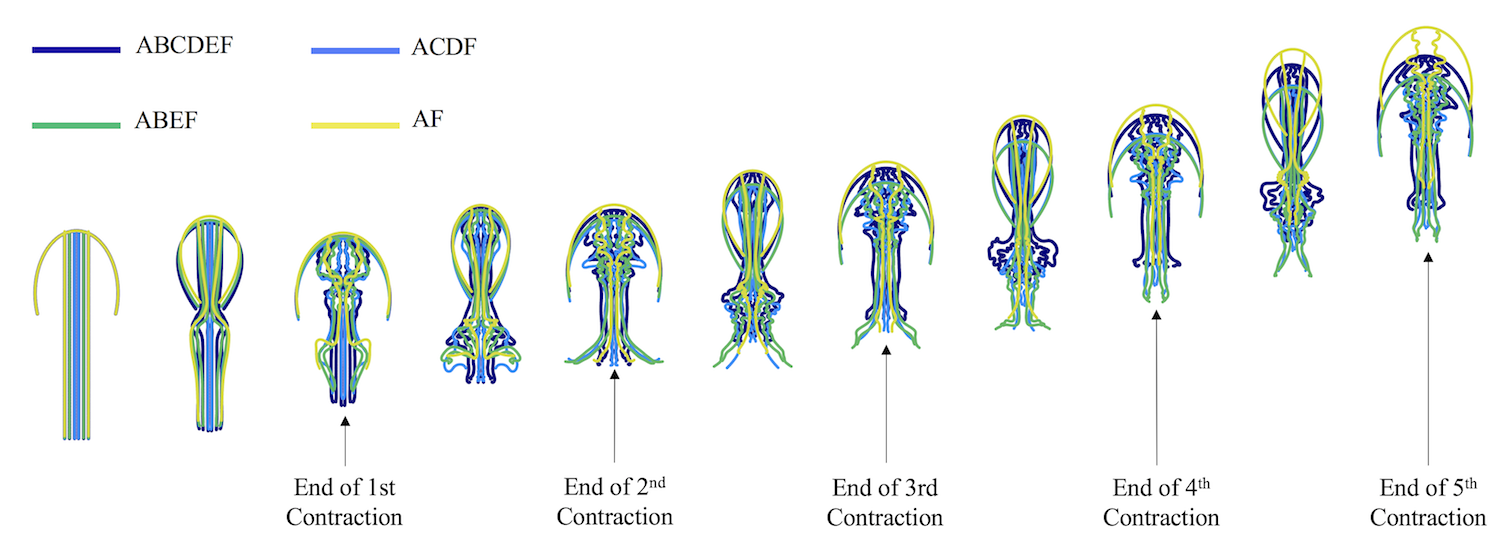}
\caption{Visualization comparing the positions of the jellyfish across the first $5$ contraction cycles for all cases considered in Section \ref{results:sec_batch1} for $Re=150$.}
\label{fig:batch1_compareLags}
\end{figure}

Qualitative analysis of forward swimming performance is given in Figure \ref{fig:batch1_compareLags}, where positions of the Lagrangian points are illustrated across the first $5$ contraction cycles for the case of $Re=150$.  It appears that the $AF$ case with only two tentacles/oral arms is able to swim faster than the other cases. Interestingly, it does not appear that less tentacles always leads to faster forward swimming as in Sections \ref{results:sec_Re} and \ref{results:sec_length}; the case of $ABCDEF$ (6 tentacles/oral arms) seems to swim slightly faster than either $ABEF$ or $ACDF$ (4 tentacle/oral arm cases); however, these were only the first $5$ contraction cycles. 

\begin{figure}[H]
\centering
\includegraphics[width=0.98\textwidth]{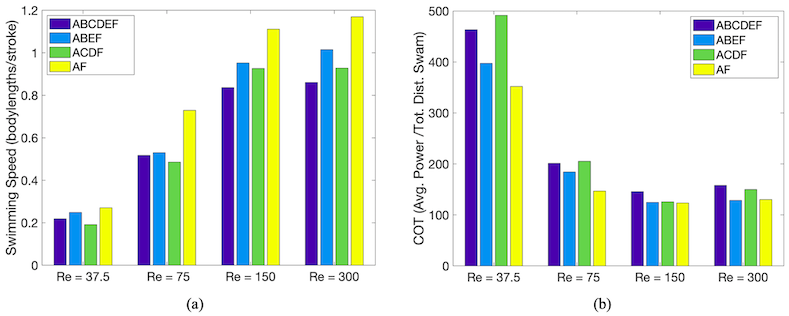}
\caption{(a) Forward swimming speed and (b) power-based cost of transport for each simulation in Section \ref{results:sec_batch1}. A nonlinear relationship between forward swimming speed, tentacle/oral arm number density and placement emerges.}
\label{fig:batch1_data}
\end{figure}

Upon computing the forward swimming speed for each case of $Re$ and tentacle/oral arm configuration considered, the case of $AF$ appears the fastest for each $Re$, see Figure \ref{fig:batch1_data}a. Figures \ref{fig:batch1_data}a and \ref{fig:batch1_data}b give the forward swimming speed and cost of transport (average power/total distance swam), respectively. For all $Re$ the $ABEF$ case was the second fastest; however, the third fastest case is the $ABCDEF$ for $Re=37.5$ and $75$, and $ACDF$ for $Re=150$ and $300$. Table \ref{table:batch1} gives the percentage decrease in swimming speeds across all cases when comparing to the case with no tentacles/oral arms. More tentacles/oral arms do not always cause lower swimming speeds, see $Re=37.5$ and $75$ where $ABCDEF$ swims faster than $ACDF$. Note that higher $Re$ generally shows a decrease in the difference between the case with no tentacles/oral arms to those with. The cost of transport is lowest in the $AF$ case across all $Re$ considered, followed by the $ABEF$ case. 


\begin{table}[h!]
\centering
\begin{center}
\begin{tabular}{ |c||c|c|c|c| } 
 \hline
        & $Re=37.5$ & $Re=75$ & $Re=150$ & $Re=300$ \\ \hline
 ABCDEF & -71.1\% & -64.1\% & -57.4\% & -57.6\% \\ 
 ABEF   & -67.5\% & -62.5\% & -49.8\% & -49.8\% \\ 
 ACDF   & -74.7\% & -65.5\% & -51.1\% & -52.6\% \\ 
 AF     & -64.6\% & -49.5\% & -42.7\% & -40.6\% \\ 
 \hline
\end{tabular}
\end{center}
\caption{Table giving the percentage decrease in forward swimming speed when compared to the case with no tentacles/oral arms.}
\label{table:batch1}
\end{table}

It is not the case that only the placement of the outermost tentacles/oral arms solely affects forward swimming speed.  There appears to be a non-linear relationship between forward swimming speed and density of tentacles/oral arms packed into the same region within a bell. We will probe this relationship further in Section \ref{results:sec_batch3}. Moreover, this section suggests that placing more tentacles/oral arms closer to the outermost ones may be beneficial for forward swimming performance; we will explore this idea further in Section \ref{results:sec_batch2}.

\begin{figure}[H]
\centering
\includegraphics[width=0.98\textwidth]{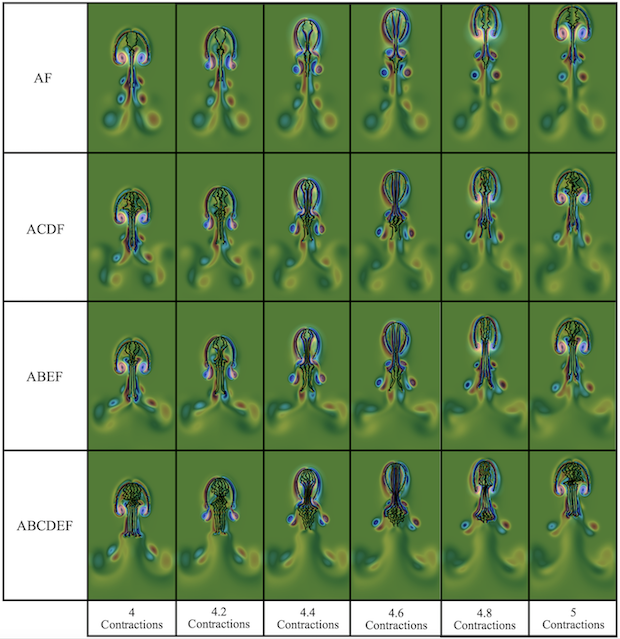}
\caption{Visualization of jellyfish position and a colormap of vorticity across the $4^{th}$ to $5^{th}$ contraction cycle for each case considered at $Re=150$.}
\label{fig:batch1_vorticity}
\end{figure}

Figure \ref{fig:batch1_vorticity} illustrates a colormap of vorticity for each different tentacle/oral arm case across the $4^{th}$ to $5^{th}$ contraction cycle for $Re=150$. In the $AF$ case, there is still a pronounced vertically-aligned vortex wake, where as in other cases of more tentacles/oral arms, the vortices begin to disperse more horizontally, as similarly suggested by previous vorticity plots (Figures \ref{fig:Compare_0_6_tents} and \ref{fig:Len_Vorticity}) as well as LCS plots (Figures \ref{fig:LCS_TentNumSweep_1T} and \ref{fig:LCS_Len_1T}). We hypothesize that the $AF$ case is the fastest because vortex formation is not as suppressed as in the other cases. However, this would not explain why the $ABCDEF$ case is faster than the $ACDF$ case for $Re\lesssim75$. This behavior may be attributed to that in the $ABCDEF$ case as there are more densely packed tentacles/oral arms, they may be acting as more of a ``rigid" wall, which allows for more elastic vortex interactions, while in the $ACDF$ cases, vortices are being ``cushioned" into the tentacles/oral arms, giving rise to more inelastic interactions, as there is more open fluid-filled space. However, for $Re\gtrsim150$, it appears this phenomena no longer suffices to provide any elastic/inelastic advantage.

%
%

\subsubsection{How does density of tentacles affect swimming performance?}
\label{results:sec_batch3}

\begin{figure}[H]
\centering
\includegraphics[width=0.6\textwidth]{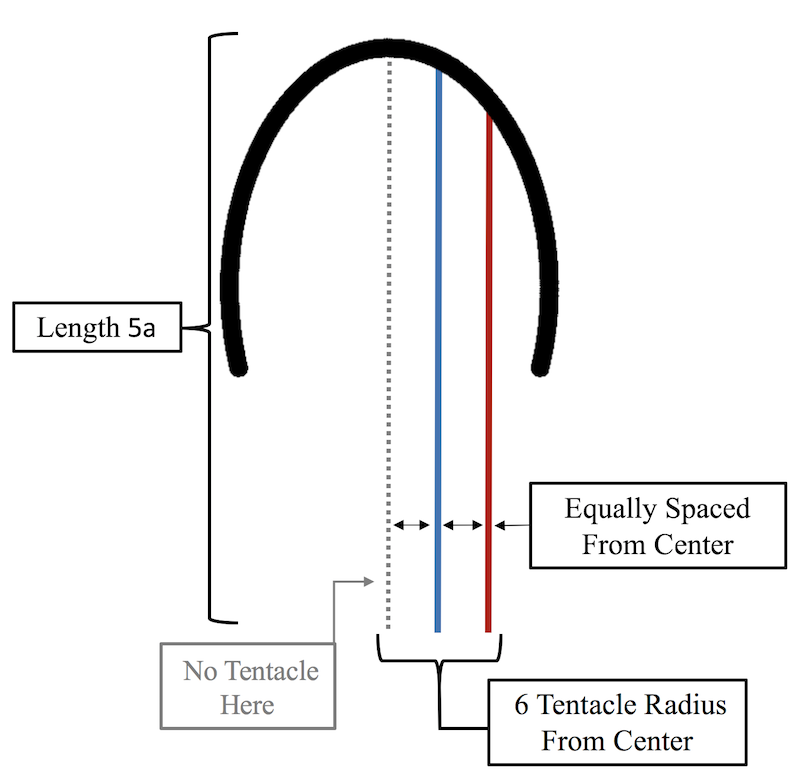}
\caption{Geometric setup for all cases considered in Section \ref{results:sec_batch3} to determine how density of the tentacles/oral arms affects forward swimming speed.}
\label{fig:batch3_geo}
\end{figure}

For this study, we will use the same placement of the outermost tentacles/oral arms as in Section \ref{results:sec_batch1} but place a different number of equally-spaced, symmetric tentacles/oral arms within the region, see Figure \ref{fig:batch3_geo}. Constructing such geometry will allow us to study effects of density of tentacles/oral arms. The even spacing was to reduce possible artifacts caused by unequal weighing of the tentacles, as in Section \ref{results:sec_batch1}'s $ABEF$ and $ACDF$ cases. We will consider four cases of differing number of tentacles/oral arms (including the outermost): (1) 2 per side  (2) 4 per side (3) 5 per side and (4) 10 per side. These numbers were chosen to keep equal spacing of the tentacles/oral arms in each subsequent case mandating that the outermost were attached to a particular Lagrangian point on the bell. Note that the previous case $ABCDEF$ in Section \ref{results:sec_batch1} (or the $6$ tentacle/oral arm case in Section \ref{results:sec_Re}) is different as the middle two tentacles/oral arms are not twice the distance apart, see Figures \ref{fig:batch1_geo} and \ref{fig:batch3_geo} for comparison. These cases were studied for $Re=37.5, 75, 150$ and $300$.  

\begin{figure}[H]
\centering
\includegraphics[width=0.98\textwidth]{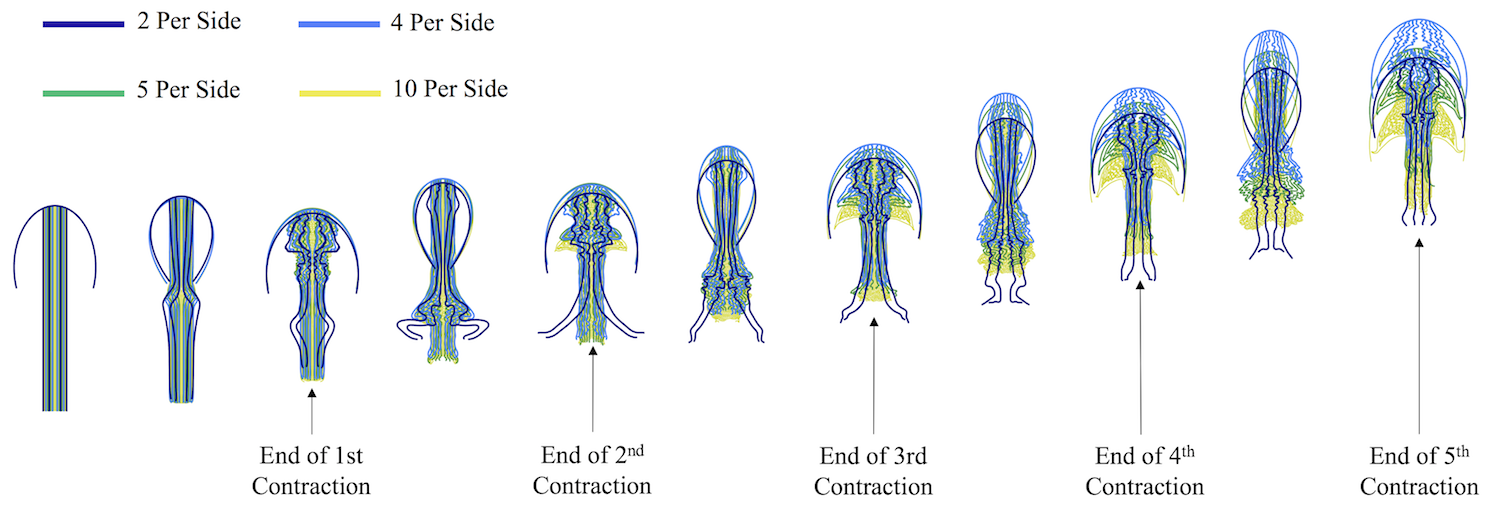}
\caption{Visualization comparing the positions of the jellyfish across the first $5$ contraction cycles for all cases considered in Section \ref{results:sec_batch3} for $Re=150$.}
\label{fig:batch3_compareLags}
\end{figure}

A qualitative analysis of forward swimming progress is given in Figure \ref{fig:batch3_compareLags}, where positions of the Lagrangian points are illustrated across the first $5$ contraction cycles for the case of $Re=150$. It appears that 4 tentacles/oral arms per side case is able to swim faster than the other cases, although this is only over the first $5$ contraction cycles. Note that less tentacles/oral arms per side do not always appear to contribute to higher swimming accelerations, as the $5$ tentacles/oral arms per side case accelerates faster than the $2$ and $10$. 

\begin{figure}[H]
\centering
\includegraphics[width=0.98\textwidth]{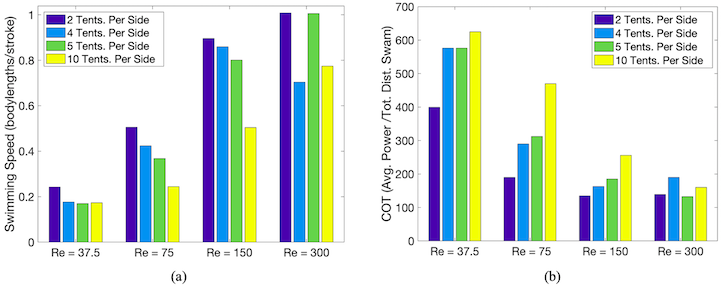}
\caption{(a) Forward swimming speed and (b) power-based cost of transport for each simulation in Section \ref{results:sec_batch3}. A nonlinear relationship between forward swimming speed, tentacle/oral arm density and placement is observed again.}
\label{fig:batch3_data}
\end{figure}

Upon quantifying swimming speeds we saw that the case with $2$ tentacles swims faster than all other cases, although barely in the $Re=300$ case (see Table \ref{table:batch3}). This data is provided in Figure \ref{fig:batch3_data}a. Only in the cases of $Re=75$ and $150$ do more tentacles always lead to decreased swimming speeds. In the $Re=37.5$ case, the $10$ tentacle per side case is slightly faster than the $5$ per side, while in the $Re=300$ case, the $5$ tentacles per side case is the second fastest (barely slower than the $2$ per side case), followed by the $10$ per side case, and finally the $4$ per side case. From Figure \ref{fig:batch3_data}b, more tentacles per side does not always attribute to greater cost of transport, e.g., the $5$ and $10$ per side case for $Re=300$.



\begin{table}[h!]
\centering
\begin{center}
\begin{tabular}{ |c||c|c|c|c| } 
 \hline
        & $Re=37.5$ & $Re=75$ & $Re=150$ & $Re=300$ \\ \hline
 2 Per Side    & -69.3\% & -65.5\% & -54.6\% &  -49.3\%  \\ 
 4 Per Side    & -77.6\% & -71.0\% & -56.4\% &  -64.6\%  \\ 
 5 Per Side    & -78.5\% & -74.9\% & -59.4\%  &  -49.5\% \\ 
 10 Per Side   & -78.1\% & -83.3\% & -74.4\% &  -61.1\%  \\ 
 \hline
\end{tabular}
\end{center}
\caption{Table giving the percentage difference in forward swimming speed when compared to the case with no tentacles/oral arms.}
\label{table:batch3}
\end{table}

To compare the cases further, we compute the relative percentage differences in swimming speeds for each respective $Re$ by comparing to case with no tentacles. This data is presented in Table \ref{table:batch3}. There is no general linear relationship between forward swimming speed and density of uniformly spaced tentacles among all cases. Such relationship only appears to manifest for $Re=75$ and $Re=150$ only; the $Re=37.5$ and $300$ cases show a nonlinear relationship. Moreover, as generally as $Re$ increases the percentage difference between the case with no tentacles and cases with tentacles/oral arms decreases here, with two exceptions - for $4$ per side between $Re=150$ and $Re=300$ and for $10$ per side between $Re=37.5$ and $Re=75$. 

\begin{figure}[H]
\centering
\includegraphics[width=0.98\textwidth]{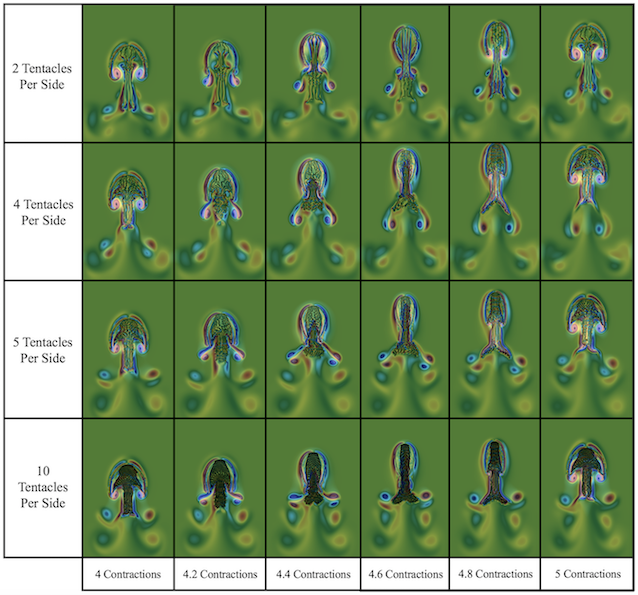}
\caption{Visualization of jellyfish position and a colormap of vorticity across its $5^{th}$ contraction cycle for each case considered at $Re=150$.}
\label{fig:batch3_vorticity}
\end{figure}

Figure \ref{fig:batch3_vorticity} gives a vorticity colormap from the $4^{th}$ to $5^{th}$ contraction cycle across each density case for $Re=150$. For every time-point shown, variations in vortex topology is observed, particularly between the $2$, $4$, and $5$ or $10$ case.  The case with $5$ or $10$ tentacles/oral arms per side show a similar vortex wake structure, while vortex wake in the case with $4$ is qualitatively different. Recall that for $Re=150$, the case with $4$ per side was the fastest swimmer. It looks as though that this configuration was able to produce the most vertically-aligned vortex wake, comparatively; however, it is not understood why \cite{Smits:2019}.

From Sections \ref{results:sec_batch1} and \ref{results:sec_batch3}, it is evident that small changes in tentacle/oral arm morphology, e.g., their placement and density, may significantly affects forward swimming speed. By observing differences among fluid scale ($Re$) in Section \ref{results:sec_batch3}, we hypothesize that fluid scale and density of tentacles/oral arms couple to vary energy (and vortex) absorption into the tentacle/oral arms during each contraction cycle, making some vortex-tentacle/oral arm interactions more elastic than others, which may enhance forward swimming speeds. This was also seen in Section \ref{results:sec_batch1} as well between cases of $ABCDEF$ and $ACDF$ for $Re\lesssim75$.

The last study we performed was biasing the placement of tentacle/oral arms towards the outermost ones. We present these results in Section \ref{results:sec_batch2}. It was motivated by the cases $ACDF$ and $ABEF$ in Section \ref{results:sec_batch1}, which showed significant differences in swimming speed for the same number of tentacles, where the case with more further placed tentacles was faster ($ABEF$).

%
%

\subsubsection{How does stacking tentacle/oral arms towards the outermost ones affect swimming performance?}
\label{results:sec_batch2}

\begin{figure}[H]
\centering
\includegraphics[width=0.6\textwidth]{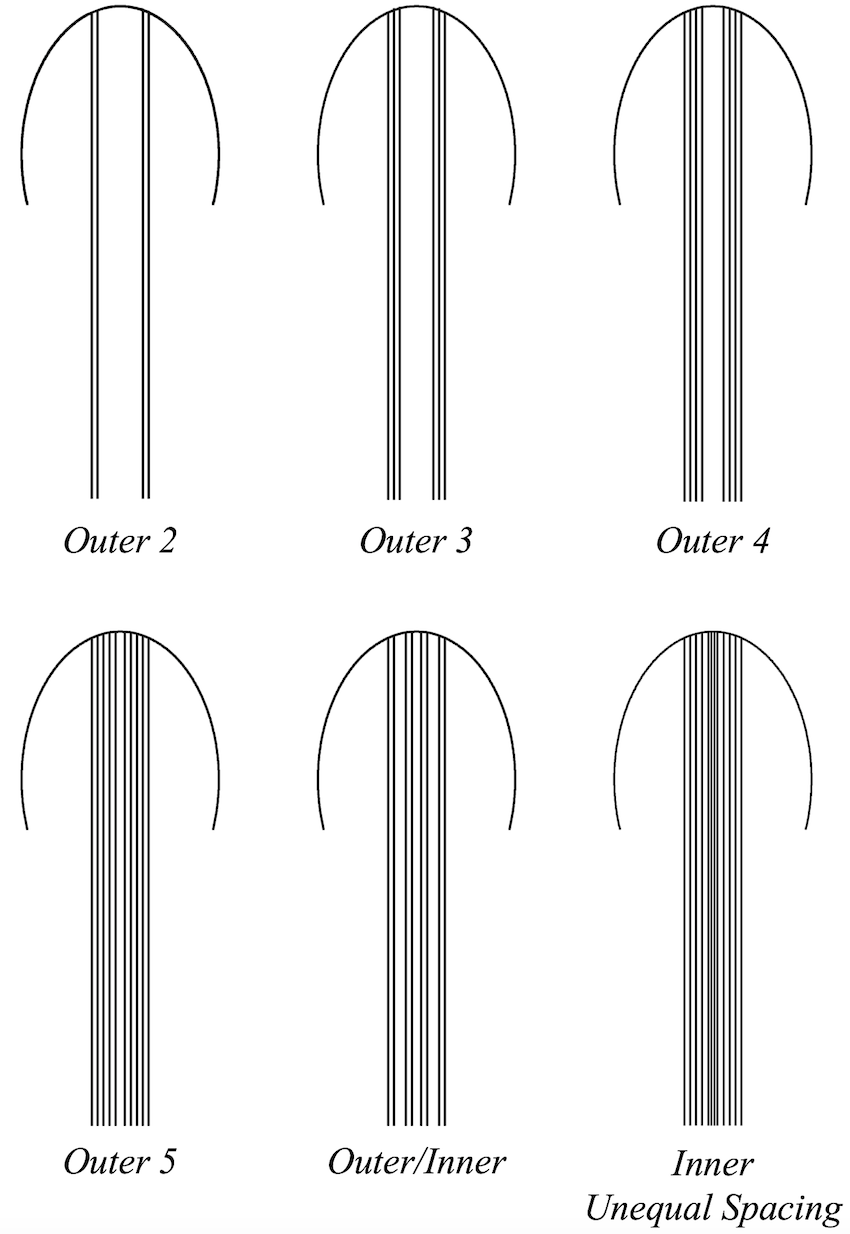}
\caption{Geometric setup for all cases considered in Section \ref{results:sec_batch2} to determine how placing more tentacles/oral arms towards the outermost ones affect forward swimming speed.}
\label{fig:batch2_geo}
\end{figure}

For this study, we will use the same placement of the outermost tentacles/oral arms as in Sections \ref{results:sec_batch1} and \ref{results:sec_batch3} but place more tentacles/oral arms towards the outermost ones, rather than equally-spacing them within the bell, as in Section \ref{results:sec_batch3}. This essentially makes clusters of the tentacles/oral arms towards the outermost ones. We will also include two other cases, not addressed in previous sections, where we include clusters of $2$ tentacles/oral arms at other locations, ``Outer/Inner", and cluster tentacles/oral arms at the midpoint, ``Inner Unequal Spacing", see Figure \ref{fig:batch2_geo} for all cases considered here. These cases were studied for $Re=37.5, 75, 150$ and $300$.  

\begin{figure}[H]
\centering
\includegraphics[width=0.95\textwidth]{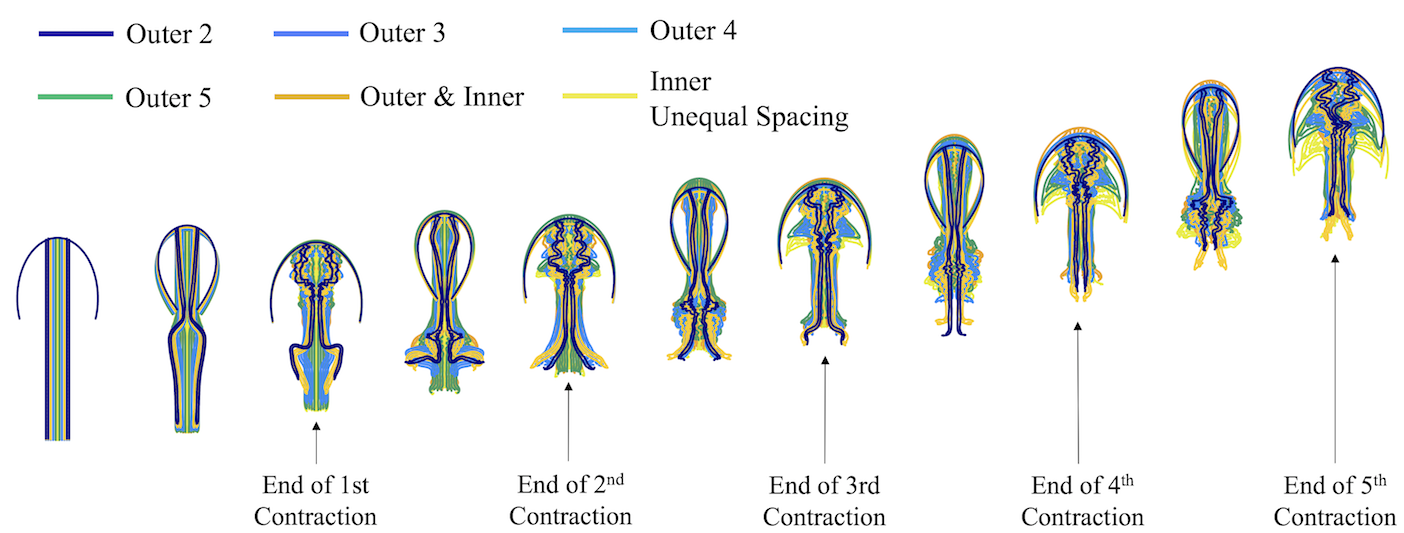}
\caption{Visualization comparing the positions of the jellyfish across the first $5$ contraction cycles for all cases considered in Section \ref{results:sec_batch2} for $Re=150$.}
\label{fig:batch2_compareLags}
\end{figure}

A qualitative analysis of forward swimming progress is given in Figure \ref{fig:batch2_compareLags}, where positions of the Lagrangian points are illustrated across the first $5$ contraction cycles in the $Re=150$ case.  The Outer 2 case looks to have swam the furthest after $5$ contraction cycles, followed closely by the Outer/Inner case and then the Outer 3 case. Note that these cases have 4, 8, and 6 tentacles/oral arms, respectively. This again illustrates that a nonlinear relationship exists between forward swimming speed and tentacle/oral arm number and density, as in Sections \ref{results:sec_batch1} and \ref{results:sec_batch3}; less tentacles/oral arms per side do not appear to always contribute to faster swimming. However, recall that during these 5 contraction cycles, steady swimming has only started to have been achieved, and so we will now quantify steady swimming speeds across the seventh and eighth contraction cycle.

\begin{figure}[H]
\centering
\includegraphics[width=0.95\textwidth]{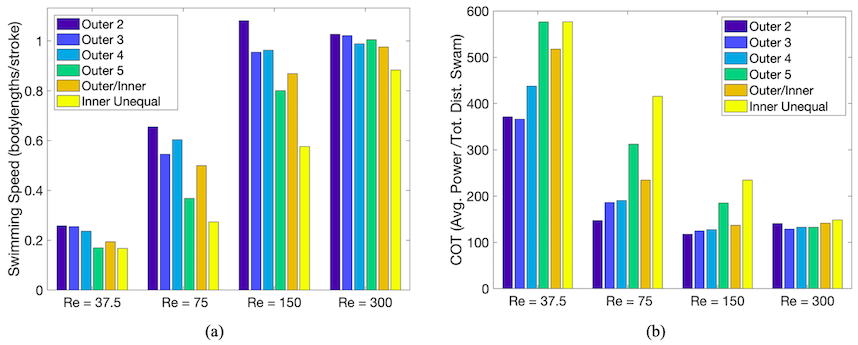}
\caption{(a) Forward swimming speed and (b) power-based cost of transport for each simulation in Section \ref{results:sec_batch2}. A nonlinear relationship between number and density of tentacles/oral arms and forward swimming speed is observed.}
\label{fig:batch2_data}
\end{figure}

Swimming speeds and cost of transport for $Re=37.5, 75, 150$, and $300$, are given in Figures \ref{fig:batch2_data}a and \ref{fig:batch2_data}b, respectively. There is a nonlinear relationship with number and density of tentacles/oral arms and forward swimming speed, even when placing them closer to the outermost ones. The Outer 2 case is the fastest case across all $Re$, although barely for $Re=37.5$ and $300$. Furthermore, the second fastest case changes between the Outer 3 and Outer 4 case; Outer 3 is second fastest for $Re=37.5$ and $300$, while Outer 4 is second fastest for $Re=75$ and $150$. For $Re=37.5$, a linear relationship between number of outer placed of tentacles/oral arms and forward swimming speed appears to emerge, though; however, it does not exist in any other cases. 

Moreover, density and placement affects swimming speed as we see that the Outer 4 and Outer/Inner cases, both which have $8$ total tentacles/oral arms (4 per side), do not swim with the same speed. Only in the $Re=300$ case do they swim at similar speeds, see Table \ref{table:batch2}, which gives the percent difference between each case described above the case of no tentacles/oral arms. Similarly to Sections \ref{results:sec_batch1} and \ref{results:sec_batch3}, percent differences generally decrease as $Re$ increases, but with two explicit exceptions - Outer 2 between $Re=150$ and $Re=300$ as well as Inner Unequal between $Re=37.5$ and $Re=75$.

\begin{table}[h!]
\centering
\begin{center}
\begin{tabular}{ |c||c|c|c|c| } 
 \hline
        & $Re=37.5$ & $Re=75$ & $Re=150$ & $Re=300$ \\ \hline
Outer 2      & -67.4\% & -55.3\% & -45.2\% &  -48.4\%  \\ 
Outer 3      & -67.7\% & -62.7\% & -51.6\% &  -48.6\%  \\ 
Outer 4      & -70.1\% & -58.8\% & -51.2\%  & -50.3\% \\ 
Outer 5      & -78.5\% & -74.9\% & -59.4\% &  -49.5\%  \\ 
Outer/Inner  & -75.4\% & -65.9\% & -55.9\%  & -51.0\% \\ 
Inner Unequal& -78.8\% & -81.3\% & -70.8\% &  -55.6\%  \\ 
 \hline
\end{tabular}
\end{center}
\caption{Table giving the percentage difference in forward swimming speed when compared to the case with no tentacles/oral arms.}
\label{table:batch2}
\end{table}

\begin{figure}[H]
\centering
\includegraphics[width=0.9\textwidth]{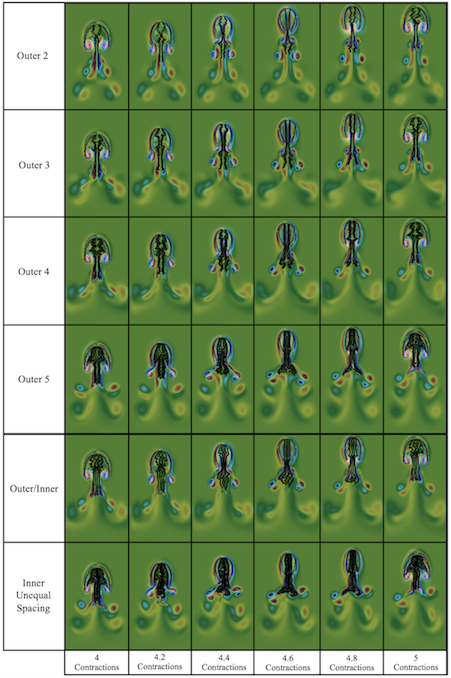}
\caption{Visualization of jellyfish position and a colormap of vorticity across its $5^{th}$ contraction cycle for each case considered at $Re=150$.}
\label{fig:batch2_vorticity}
\end{figure}

Similar to Sections \ref{results:sec_batch1} and \ref{results:sec_batch3}, Section \ref{results:sec_batch2} highlights the existence of a complex relationship between fluid scale ($Re$), placement, number, and density of tentacles/oral arms in regards to potential forward swimming performance. Different variations of these parameters give rise to differing vortex wakes (see Figure \ref{fig:batch2_vorticity}), which could possibly be used to predict enhanced or inhibited swimming performance \cite{Miles:2019}; however, such relationships are nontrivial or may be impossible to discern accurately \cite{Smits:2019,Floryan:2019}. In particular, different fluid scales ($Re$) and tentacle/oral arm densities may determine whether the tentacles/oral arms act as energy absorbing entities or otherwise (elastic/inelastic vortex collisions). There are further intricate relationships to decrypt on how this either enhances or inhibits forward swimming. 

%
%

%
%

\section{Discussion and Conclusion}

Previous fluid-structure interaction models have shown that a jellyfish's bell morphology, material properties, and kinematics affect its potential forward swimming speeds \cite{Hoover:2015,Hoover:2017,Miles:2019}. This is the first computational study to reveal that including tentacles/oral arms will also significantly affect forward swimming performance in jellyfish. Using an idealized $2D$ computational model, we illustrated a complex relationship between tentacle/oral arm morphology (number, length, placement, and density) and fluid scale ($Re$) on forward swimming speed. 

In particular we discovered that including more symmetrically and equally spaced tentacles/oral arms within its bell, will greatly inhibit forward swimming speed (Section \ref{results:sec_Re}). Similarly to previous studies \cite{Hershlag:2011,Miles:2019}, forward swimming speed steadies out as $Re$ increases, even in cases with tentacles/oral arms; the tentacles/oral arms simply lowers the maximal swimming forward speed achievable (Section \ref{results:sec_Re}). K. Katija in 2015 \cite{Katija:2015} saw a similar percentage decrease in swimming speed for Australian Spotted Jellyfish with and without oral arms. Moreover, including more symmetrically/equally spaced tentacles/oral arms seems to increase horizontal mixing by the jellyfish bell, which decreases vertical mixing as well. The vertical-size of the vortex wake also decreases. Thus jellyfish morphology not only can limits its intrinsic swimming ability, but overall fluid transport and mixing \cite{Costello:2008,Katijai:2015}.  

For $Re=150$, which is the approximate scale of a \textit{Sarsia tubulosaa}, varying the length of the tentacles/oral arms showed the existence of three possible swimming performance states. (1) If their length is short enough, forward swimming speeds are not significantly different than the case of no tentacles/oral arms. (2) For long tentacles/oral arms, swimming speed asymptotically steadies out, and longer tentacles/oral arms will not significantly affect forward swimming speeds. (3) There is a middle range in which swimming speed drops off from approximately the case with no tentacles/oral arms to where swimming speed asymptotically steadies out. It does not appear that longer tentacles will drop forward swimming speed to zero. Tentacles/oral arms of varying lengths lengths have different positional dynamics change during a contraction, where short enough tentacles/oral arms are pulled upwards into the bell with minimal amounts exposed out of the bell, while long enough tentacles are pulled upwards into the bell as well, but with the majority of the tentacle/oral arm still hanging outside the bell (Section \ref{results:sec_length}).  Thus, as tentacles/oral arms get longer, more fluid mixing occurs near the jellyfish bell, rather than downstream in the vortex wake.

A nonlinear relationship between tentacle/oral arm placement and density and swimming speed was also observed (Section \ref{results:sec_density}). It is not that case that only placement of the outermost tentacles/oral arms dictates the inhibition of forward swimming speed (Section \ref{results:sec_batch1}). Less dense tentacle/oral arm configurations (within the same region) did not always lead to increased swimming performance (Section \ref{results:sec_batch3}). Stacking more tentacles/oral arms closer to the outermost ones, did not always lead to lower forward swimming speeds (Section \ref{results:sec_batch2}). Furthermore, small changes in placement of the same number of tentacles/oral arms may results in non-linear differences in swimming speeds (Sections \ref{results:sec_batch1} and \ref{results:sec_batch2}).

This study showed that small changes in tentacle/oral number, placement, density, or length significantly affected forward swimming performance in an $2D$ idealized jellyfish model. Any effects of tentacle/oral arm stiffness were not thoroughly investigated. Investigating the effect of tentacle/oral arms for a variety of bell morphologies and kinematics would provide further insight into jellyfish ecology. In particular, modeling a specific jellyfish species' bell and tentacle/oral morphology and bell kinematics could highlight why it uses a particular predation strategy over others. 

Even certain jellyfish species (\textit{Turritopsis nutricula}) that have been deemed to have the potential for immortality \cite{Piraino:1996}, need to eat to sustain themselves. Under an evolutionary lens, locomotive mechanisms are largely believed to be based on the size of an organism \cite{Vogel:1996}. Jellyfish have evolved and adapted to use a diverse variety of foraging strategies across a variety of sizes and morphologies. Some are active hunters, such as the sea wasp (box jellyfish, \textit{Chironex fleckeri}), while others are more opportunist predators, who passively drift and wait for prey, such as the Lions Mane jellyfish (\textit{Cyanea capillata}). Both species have significantly different bell, tentacle, oral arm morphologies, bell kinematics, as well as, forward swimming speeds \cite{Purcell:2001,Purcell:2003,BastianThesis:2011,Colin:2013,Crawford:2016}. Therefore, a foraging advantage for the Lions Mane jellyfish are its longer, numerous, and dense tentacles, as it does not actively hunt, possibly due to a constraint between such morphology and forward swimming performance. On the other hand, the box jellyfish may not benefit from more dense or numerous tentacles, as it could potentially inhibit its swimming performance and thereby its predation strategy. The evolution of jellyfish may detail an interesting story between shape and size (bell morphology, tentacle/oral arm number, placement, density, and length) and function (active hunting or passive foraging strategies), where evolutionary bifurcations and adaptations gave rise to some of the oldest (most successful) but possibly laziest (passive, drift eating), or fear-inducing (active hunting), efficient organisms.

\vspace{6pt} 



%
%

\authorcontributions{Conceptualization, J.G.M. and N.A.B.; Data curation, N.A.B.; Formal analysis, N.A.B.; Funding acquisition, J.G.M. and N.A.B.; Investigation, J.G.M. and N.A.B.; Methodology, J.G.M. and N.A.B.; Project administration, N.A.B.; Resources, N.A.B.; Software, N.A.B.; Supervision, N.A.B.; Validation, N.A.B.; Visualization, J.G.M. and N.A.B.; Writing - original draft, J.G.M. and N.A.B.; Writing - review \& editing, J.G.M. and N.A.B..}

%
%

\funding{J.G.M. was partially funded by the Bonner Community Scholars Program and Innovative Projects in Computational Science Program (NSF DUE \#1356235) at TCNJ. N.A.B. was funded and supported the NSF OAC-1828163, TCNJ Support of Scholarly Activity (SOSA) Grant, the Department of Mathematics and Statistics, and the School of Science at TCNJ.}

%
%

\acknowledgments{The authors would like to thank Laura Miller and Alexander Hoover for sharing their knowledge and passion of jellyfish locomotion and Yoshiko Battista for introducing N.A.B. to the world of marine life. We would also like to thank Christina Battista, Robert Booth, Christina Hamlet, Matthew Mizuhara, Arvind Santhanakrishnan, Emily Slesinger, and Lindsay Waldrop for comments and discussion.}

\conflictsofinterest{The authors declare no conflict of interest.} 

\abbreviations{The following abbreviations are used in this manuscript:\\

\noindent 
\begin{tabular}{@{}ll}
$Re$ & Reynolds Number \\
$IB$ & Immersed Boundary Method \\
$COT$ & Cost of Transport 
\end{tabular}}

%
%

\appendixtitles{yes} 
\appendixsections{multiple} 

\appendix
%
%

\section{Details on IB}
\label{IB_Appendix}

A two-dimensional formulation of the immersed boundary (IB) method is discussed below. The IB software used was \textit{IB2d} \cite{Battista:2015,BattistaIB2d:2017,BattistaIB2d:2018}. The software has been validated \cite{BattistaIB2d:2017} with specific convergence tests performed for the jellyfish model in \cite{Battista:2019} and \cite{Miles:2019}. For a full review of the immersed boundary method, please see Peskin 2002 \cite{Peskin:2002} or Mittal et al. 2005 \cite{Mittal:2005}. 

%
%

\subsection{Governing Equations of IB}

The conservation of momentum and mass equations that govern an incompressible and viscous fluid are listed below:

\begin{equation} 
   \rho\Big[\frac{\partial\U}{\partial t}({\bf x},t) +\U({\bf x},t)\cdot\nabla \U({\bf x},t)\Big]=  \nabla p({\bf x},t) + \mu \Delta \U({\bf x},t) + \F({\bf x},t) \label{eq:NS1}
 \end{equation}
  \begin{equation}
      \div \U({\bf x},t) = 0 \label{eq:NSDiv1}
  \end{equation}
where $\U({\bf x},t) $ is the fluid velocity, $p({\bf x},t) $ is the pressure, $\F({\bf x},t) $ is the force per unit area applied to the fluid by the immersed boundary, $\rho$ and $\mu$ are the fluid's density and dynamic viscosity, respectively. The independent variables are the time $t$ and the position ${\bf x}$. The variables $\U, p$, and $\F$ are all written in an Eulerian frame on the fixed Cartesian mesh, $\textbf{x}$. 

The interaction equations, which handle all communication between the fluid (Eulerian) grid and immersed boundary (Lagrangian grid) are the following two integral equations:
\begin{align}
   {\bf F}({\bf x},t) &= \int {\bf f}(s,t)  \delta\left({\bf x} - {\bf X}(s,t)\right) dq \label{eq:force1} \\
   {\bf U}({\bf X}(s,t))  &= \int \U({\bf x},t)  \delta\left({\bf x} - {\bf X}(s,t)\right) d{\bf x} \label{eq:force2}
\end{align}
where ${\bf f}(s,t)$ is the force per unit length applied by the boundary to the fluid as a function of Lagrangian position, $s$, and time, $t$, $\delta({\bf x})$ is a three-dimensional delta function, and ${\bf X}(s,t)$ gives the Cartesian coordinates at time $t$ of the material point labeled by the Lagrangian parameter, $s$. The Lagrangian forcing term, ${\bf f}(s,t)$, gives the deformation forces along the boundary at the Lagrangian parameter, $s$. Equation (\ref{eq:force1}) applies this force from the immersed boundary to the fluid through the external forcing term in Equation (\ref{eq:NS1}). Equation (\ref{eq:force2}) moves the boundary at the local fluid velocity. This enforces the no-slip condition. Each integral transformation uses a three-dimensional Dirac delta function kernel, $\delta$, to convert Lagrangian variables to Eulerian variables and vice versa.


Using delta functions as the kernel in Eqs.(\ref{eq:force1}-\ref{eq:force2}) is what gives IB its power. To approximate these integrals, discretized (and regularized) delta functions are used. We use the ones given from \cite{Peskin:2002}, e.g., $\delta_h(\mathbf{x})$, 
\begin{equation}
\label{delta_h} \delta_h(\mathbf{x}) = \frac{1}{h^3} \phi\left(\frac{x}{h}\right) \phi\left(\frac{y}{h}\right) \phi\left(\frac{z}{h}\right) ,
\end{equation}
where $\phi(r)$ is defined as
\begin{equation}
\label{delta_phi} \phi(r) = \left\{ \begin{array}{l} \frac{1}{8}(3-2|r|+\sqrt{1+4|r|-4r^2} ), \ \ \ 0\leq |r| < 1 \\    
\frac{1}{8}(5-2|r|+\sqrt{-7+12|r|-4r^2}), 1\leq|r|<2 \\
0 \hspace{2.1in} 2\leq |r|.\\
\end{array}\right.
\end{equation}

%
%

\subsection{Numerical Algorithm}
As stated in the main text, we impose periodic and no slip boundary conditions on the rectangular domain. To solve Equations (\ref{eq:NS1}), (\ref{eq:NSDiv1}),(\ref{eq:force1}) and (\ref{eq:force2}) we need to update the velocity, pressure, position of the boundary, and force acting on the boundary at time $n+1$ using data from time $n$. The IB does this in the following steps \cite{Peskin:2002}.

\textbf{Step 1:} Find the force density, ${\bf{F}}^{n}$ on the immersed boundary, from the current boundary configuration, ${\bf{X}}^{n}$.\\
\indent\textbf{Step 2:} Use Equation (\ref{eq:force1}) to spread this boundary force from the Lagrangian boundary mesh to the Eulerian fluid lattice points.\\
\indent\textbf{Step 3:} Solve the Navier-Stokes equations, Equations (\ref{eq:NS1}) and (\ref{eq:NSDiv1}), on the Eulerian grid. Upon doing so, we are updating ${\bf{u}}^{n+1}$ and $p^{n+1}$ from ${\bf{u}}^{n}$, $p^{n}$, and ${\bf{f}}^{n}$. Note that a staggered grid projection scheme is used to perform this update.\\
\indent\textbf{Step 4:} Update the material positions, ${\bf{X}}^{n+1}$,  using the local fluid velocities, ${\bf{U}}^{n+1}$, computed from ${\bf{u}}^{n+1}$ and Equation (\ref{eq:force2}).
%

%
%

\section{Varying the Poroelastic Coefficient, $\alpha$}
\label{app:varying_alpha}

\begin{figure}[H]
\centering
\includegraphics[width=0.6\textwidth]{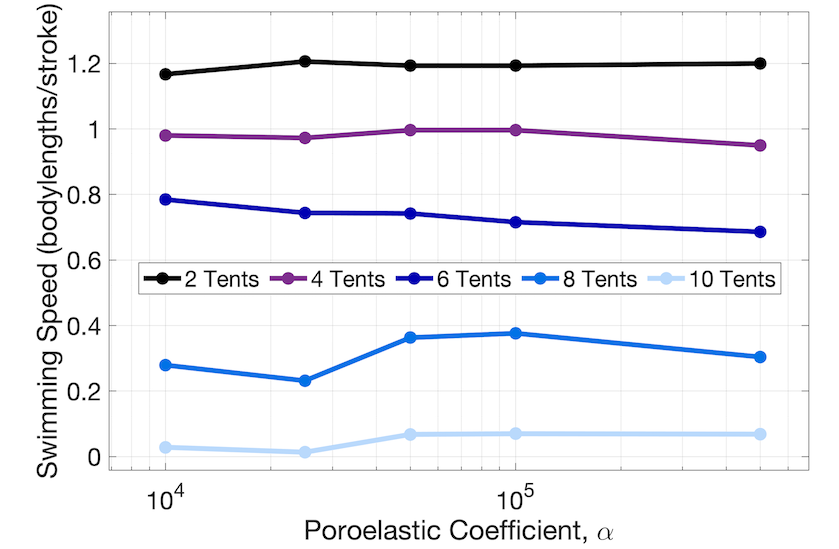}
\caption{Illustrating forward swimming speeds for different Poroelasticity Coefficients, $\alpha$, at $Re=150$.}
\label{fig:Poro_Speed}
\end{figure}

Varying $\alpha$ did not significantly affect forward swimming speeds for different numbers of tentacles/oral arms. Figure \ref{fig:Poro_Speed} shows that for $Re=150$ varying $\alpha$ did not drastically affect forward swimming speeds. Numerical stability issues were encountered for $\alpha<10^4$. The scope of this work was to explore general inhibitions on swimming performance by the addition of tentacles/oral arms, rather than focusing on effects of different poroelasticities.

%
%

\section{Varying the Reynolds Number, $Re$}
\label{app:varying_Re}

\begin{figure}[H]
\centering
\includegraphics[width=0.6\textwidth]{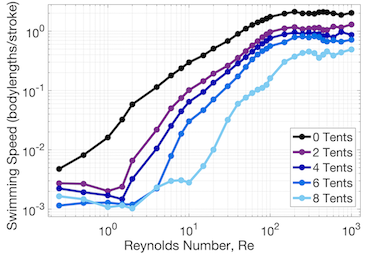}
\caption{Illustrating forward swimming speeds for a spectrum of $Re$ and numbers of tentacles.}
\label{fig:ReSpeedLogLog}
\end{figure}

This data was previously presented in Section \ref{results:sec_Re}; however, here we present the data in logarithmic form to more clearly illustrate swimming speeds at lower $Re$, see Figure \ref{fig:ReSpeedLogLog}. Figure \ref{fig:ReSpeedLogLog} illustrates that over a particular range of $Re$ that swimming speed geometrically increases for a specified tentacle number. Interestingly, the geometric increase appears uniform across every case of differing tentacle number, as the linear slope on this logarithmic plot is approximately the same. 

\begin{figure}[H]
\centering
\includegraphics[width=0.975\textwidth]{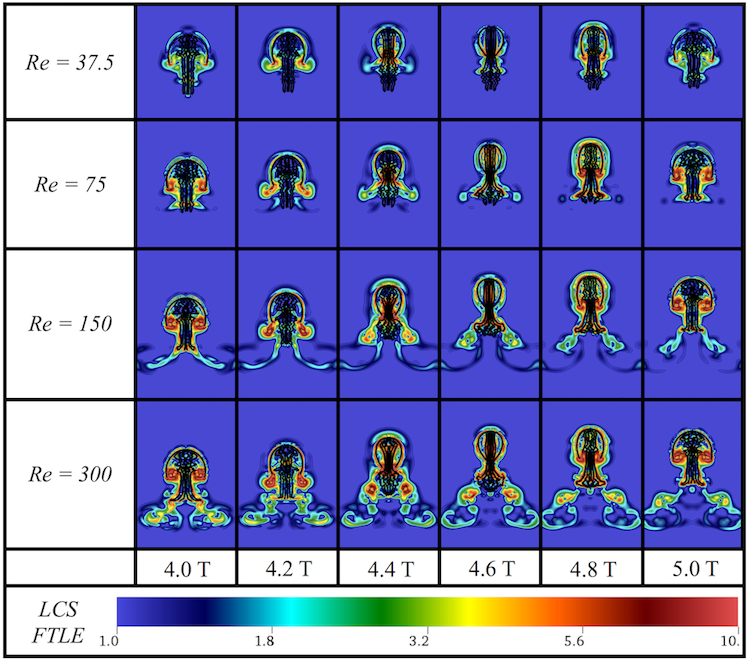}
\caption{Visualization comparing Lagrangian Coherent Structures (LCS) using finite-time Lyanpunov exponents (FTLE) for the case with 6 total tentacles/oral arms (3 symmetrically placed per side) and $Re=\{37.5,75,150,300\}$ between the $4^{th}$ and $5^{th}$ contraction cycle.}
\label{fig:LCS_Re}
\end{figure}

Figure \ref{fig:LCS_Re} compares the FTLE LCS analysis over one contraction cycle (between the $4^{th}$ and $5^{th}$) between $Re=\{37.5,75,150,300\}$ for the case of $6$ total tentacles/oral arms (3 symmetrically placed on each side of the bell).

\begin{figure}[H]
\centering
\includegraphics[width=0.975\textwidth]{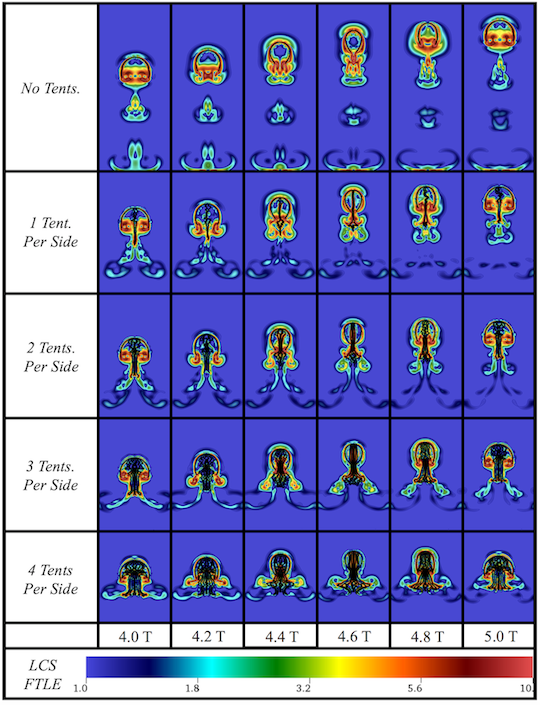}
\caption{Visualization comparing Lagrangian Coherent Structures (LCS) using finite-time Lyanpunov exponents (FTLE) for cases with either $0,1,2,3$ or $4$ symmetrically-placed tentacles/oral arms per side for $Re=150$ between the $4^{th}$ and $5^{th}$ contraction cycle.}
\label{fig:LCS_TentNumSweep}
\end{figure}

Figure \ref{fig:LCS_TentNumSweep} compares the FTLE LCS analysis over one contraction cycle (between the $4^{th}$ and $5^{th}$) between cases of differing number of tentacles at $Re=150$. 

%
%

\section{Varying the Tentacle/Oral Arm Length}
\label{app:varying_length}

\begin{figure}[H]
\centering
\includegraphics[width=0.975\textwidth]{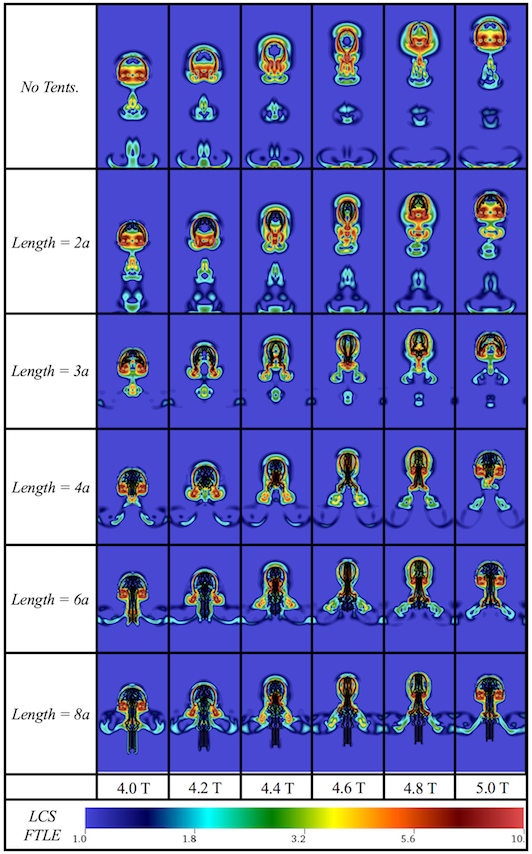}
\caption{Visualization comparing Lagrangian Coherent Structures (LCS) using finite-time Lyanpunov exponents (FTLE) for the case with 6 total tentacles/oral arms (3 symmetrically placed per side) of varying lengths (in multiples of the bell radius, $a$, between the $4^{th}$ and $5^{th}$ contraction cycle.}
\label{fig:LCS_Len}
\end{figure}

Figure \ref{fig:LCS_Len} compares the FTLE LCS analysis over one contraction cycle (between the $4^{th}$ and $5^{th}$) between cases of differing tentacle/oral arm lengths at $Re=150$. 

\reftitle{References}

\bibliographystyle{mdpi} 
\bibliography{jelly}   

\end{document}